\documentclass[a4paper]{article}

\usepackage[utf8]{inputenc}

\usepackage{graphicx,color}

\usepackage{amsfonts} % pour le N des naturels
\usepackage{amssymb}

\newcommand\set[1]{\ensuremath{\{ #1 \} }}
\newcommand\inter[1]{\ensuremath{[#1]}}
\newcommand\cur[1]{\ensuremath{{\cal{#1}}}}

\newtheorem{definition}{Definition}
\newtheorem{remark}{Remark}

\begin{document}

\date{}

\title{Non-Altering Time Scales for Aggregation of Dynamic Networks into Series of Graphs\thanks{This is a complete version of the extended abstract appeared in CoNEXT 2015~\cite{LCF15}.}}

\author{%%%% First author details
{\sc Yannick L{\'e}o},\\[2pt]
{\small Univ Lyon, ENS de Lyon, CNRS, Inria, UCB Lyon 1, LIP UMR 5668,}\\{\small 15 parvis Ren{\'e} Descartes, F-69342, Lyon, FRANCE}\\
 \texttt{yannick.leo@inria.fr}\\[6pt]
%%%%%%% Second author details
{\sc Christophe Crespelle}\\[2pt]
{\small Univ Lyon, UCB Lyon 1, CNRS, ENS de Lyon, Inria, LIP UMR 5668,}\\{\small 15 parvis Ren{\'e} Descartes, F-69342, Lyon, FRANCE and}\\{\small Institute of Mathematics, Vietnam Academy of Science and Technology,}\\{\small 18 Hoang Quoc Viet, Hanoi, Vietnam}\\
\texttt{christophe.crespelle@inria.fr}\\[6pt]
%%%%%%%
{\sc and}\\[6pt]
%%%%%%% Third author details
{\sc Eric Fleury} \\[2pt]
{\small Univ Lyon, ENS de Lyon, Inria, CNRS, UCB Lyon 1, LIP UMR 5668,}\\{\small 15 parvis Ren{\'e} Descartes, F-69342, Lyon, FRANCE}\\
 \texttt{eric.fleury@inria.fr}
}

\maketitle

\begin{abstract}
Many dynamic networks coming from real-world contexts are \emph{link streams}, i.e. a finite collection of triplets $(u,v,t)$ where $u$ and $v$ are two nodes having a link between them at time $t$. A very large number of studies on these objects start by aggregating the data in disjoint time windows of length $\Delta$ in order to obtain a series of graphs on which are made all subsequent analyses. Here we are concerned with the impact of the chosen $\Delta$ on the obtained graph series. We address the fundamental question of knowing whether a series of graphs formed using a given $\Delta$ faithfully describes the original link stream. We answer the question by showing that such dynamic networks exhibit a threshold for $\Delta$, which we call the \emph{saturation scale}, beyond which the properties of propagation of the link stream are altered, while they are mostly preserved before. We design an automatic method to determine the saturation scale of any link stream, which we apply and validate on several real-world datasets.
%{dynamic network, aggregation, time scale, link stream, graph series, occupancy.}
%%%% If classification number provided then
%\\
%2000 Math Subject Classification: 34K30, 35K57, 35Q80,  92D25
\end{abstract}

%\marginpar{chris: double affiliation, remerciements CNRS region rhone-alpes}

%%%%%%%%%%%%%%%%%%%%%%%%%%%%%%%%%%%%%%%%%%%%%%%%%
%%%%%%%%%%%%%%%%%%%%%%%%%%%%%%%%%%%%%%%%%%%%%%%%%
\section{Introduction}
%%%%%%%%%%%%%%%%%%%%%%%%%%%%%%%%%%%%%%%%%%%%%%%%%
%%%%%%%%%%%%%%%%%%%%%%%%%%%%%%%%%%%%%%%%%%%%%%%%%

Many real world dynamic networks are naturally given in the form of a finite collection $\cur{L}$ of triplets $(u,v,t)$, which we call a \emph{link stream}, where $u,v\in V$ are two nodes of the network and $t$ is a timestamp\footnote{Time can be continuous or discrete. The method we design works in both frameworks. Though, the sample datasets on which we illustrate it all use discrete timestamps.}, with the meaning that nodes $u$ and $v$ have a link between them at time $t$. Depending on the context\footnote{See~\cite{holme2012temporal} for a survey on dynamic networks and the vocabulary used to refer to them in the various contexts in which they are encountered and studied.}, these links can represent physical contacts between individuals, exchanges of emails between people, commercial interactions between companies, etc.
When one wants to study such dynamic networks, a very common approach~\cite{EP06,LMS+08,MCF14,HMK+07,CZS+07,SPY+07,DS04,PGP+12,NCF12,
VMC+09,VBC+13,SM05,STF06,PC15,CGW+08,LKF07,GBB09,BB06,LGK10} is to transform them into series of graphs% $G_{t'}$, where $t'\in \inter{1,T'}$
. The process used to do so is called \emph{aggregation}. It consists in choosing a time window $[a,b]\subseteq\inter{0,T}$ in the initial series, where $T$ is the length of the period of study, and forming the graph $G_{[a,b]}$ with all edges $u,v$ such that there exists a triplet $(u,v,t)\in \cur{L}$ with $t\in[a,b]$. Doing so for a collection of windows that covers the entire period of study, one obtains a representation of the dynamic network as a graph series: the graphs formed for each window, called \emph{snapshots}. Very often, as in this paper, the windows are disjoint and all have the same length, but in some studies, they may also overlap~\cite{LHM13,BM06,MMB05,SV+11,CV+10,SBK12} or have different lengths~\cite{RFM+09,STK+16} or all start at the beginning of the period of study~\cite{LM08,OC04,Hol03,SBK12}. In all cases, once the series is obtained, all analyses are conducted on it instead of the raw data, which is in the form of a link stream, referred to as the \emph{original link stream} in the rest of the paper, see Figure~\ref{fig:ex-agreg}.

There are two main reasons to use aggregation for studying dynamic networks. First, in many cases, it does not make sense to study the network at the scale of the time resolution of the timestamps of the given link stream. For example, in an email dataset, the timestamps of the events (sending of email) are often given with a 1-second resolution. However, studying the dynamic network at this time scale does not give a general and comprehensive view of its organization, like someone watching a painting with a microscope. Hence, aggregation allows to study the network at a scale which is relevant compared to its activity.
The second reason for using aggregation is that it produces graphs. They give an instantaneous view of the network (snapshot) which is practical in itself to get a view of what the object under study looks like and one can use the rich set of notions developed in graph theory to analyze the considered dynamic network.

%Aggregation offers some clear benefits: i) changing the scale of study can give a more relevant view of the dynamics and ii) obtaining graphs allows to use the rich set of notions developed in graph theory to analyze the dynamic network.
%If the benefits of aggregation are clear, on the other hand, it also raises some important concerns, as the length chosen for the aggregation window has a strong impact on the properties of the aggregated graph series and therefore on the conclusions derived from analysis.

If the benefits of aggregation are clear, on the other hand, it also raises some important concerns. Indeed, the length chosen for the aggregation window usually has a strong impact on the properties of the aggregated graph series~\cite{KKB+12,RPB13,SBK12}. This raises the question of which time scale should be chosen to study a given dynamic network and how much the properties studied, based on which conclusions are derived, are sensitive to the length of the aggregation period used \cite{PGP+12,NCF12,VL14}. It points out that in any case, this period should not be chosen without well established evidence, as it is currently done in most of the studies cited above.
Pushing further, it is not even clear whether an aggregated series faithfully describes the original link stream. Indeed, the aggregation process goes along with a loss of information: in each aggregation window, the information on the exact times at which links occur in this window is lost. In particular, in a given time window, it is impossible to know whether a given link $(a,b)$ has occurred before or after another one $(b,c)$. This question, which determines whether it is possible to go from node $a$ to node $c$, via $b$, within this time window (only if $ab$ has occurred before $bc$), is crucial for many phenomena taking place on the dynamic network, such as epidemic spreads, possibilities of communications and cascade of influence for example. The wider the aggregation period, the greater the amount of information lost. At the limit, aggregating a link stream over the whole period of study yields one single static network which misses all the information on the order of occurrences of links and which therefore very poorly captures the structure of the original dynamic network, see e.g. \cite{Moo02}. Then, more generally, for a given aggregation period, one can ask whether the obtained graph series is a faithful representation of the original link stream. This is precisely the question we address here, through the prism of propagation properties.%\\

%\marginpar{refs pour agreg totale = pas bien?}
% y a au moins J. Moody 2002, et refs therein?

%\vspace*{0.5em}
%\smallskip

%An even more crucial issue is the relationship between the aggregated graph series and the original network, given in the form of a temporal series of links. 

%Then, there is a balance to find between an aggregation period large enough to obtain a relevant view of the dynamic network and small enough so that the main part of the temporal information of the dynamic network is preserved when forming the graph series. Ideally, one would like to determine the larger possible aggregation period resulting in a reasonable loss of information. This is exactly the meaning of the aggregation period $\gamma$ returned by our method. The network should not be studied with an aggregation period significantly greater than $\gamma$ but rather with aggregation periods lower than $\gamma$ and approximately of the same order of magnitude as $\gamma$. 

%%%%%%%%%%%%%%%%%%%%%%%%%%%%%%%%%%%%%%%%%%%%%%%%%
%%%%%%%%%%%%%%%%%%%%%%%%%%%%%%%%%%%%%%%%%%%%%%%%%
\subsection{Our results}
%%%%%%%%%%%%%%%%%%%%%%%%%%%%%%%%%%%%%%%%%%%%%%%%%
%%%%%%%%%%%%%%%%%%%%%%%%%%%%%%%%%%%%%%%%%%%%%%%%%

%\noindent\emph{Our results.}
We show that for many dynamic networks, the length $\Delta$ of the window chosen for aggregating the network into a graph series exhibits a threshold, which is proper to each network. Beyond this threshold, the propagation properties of the graph series obtained from aggregation show evidence of alteration, while they are mostly preserved below it. We design a method, called the \emph{occupancy method}, in order to determine this threshold, which we call the \emph{saturation scale} and denote $\gamma$. We apply and validate the occupancy method on various real-world datasets, as well as on synthetic dynamic networks.

This answers the fundamental question of deciding whether a given aggregation period gives rise to a graph series that faithfully describes the original dynamic network. The aggregation periods beyond the saturation scale alters the properties of propagation of the dynamics. This range of scale must then be avoided or used only for analyzing properties of the series that do not suffer this alteration. 

Moreover, the saturation scale, which is the larger non-altering aggregation period, is a characteristic time scale of the network. It can then be used to compare the properties of different dynamic networks at a same level of aggregation, which is very interesting in itself. Finally, let us emphasize that our method is fully automatic and does not require any parameter as input. Therefore, it can easily been incorporated into any automatic tool for analyzing dynamic networks.

%Here we use this property to lead extensive simulations and determine the dependency of the saturation scale on several parameters of the networks.

%Here, we design an automatic method to determine $\gamma$, which has the remarkable feature to be self-certifying: it also provides an estimation of the accuracy of the returned value of $\gamma$.
%And we show that our method can also be used to determine whether sensor-measured dynamic networks have been measured with a sufficiently fine sampling period.

%\vspace*{0.5em}

%%%%%%%%%%%%%%%%%%%%%%%%%%%%%%%%%%%%%%%%%%%%%%%%%
%%%%%%%%%%%%%%%%%%%%%%%%%%%%%%%%%%%%%%%%%%%%%%%%%
\subsection{Related work}
%%%%%%%%%%%%%%%%%%%%%%%%%%%%%%%%%%%%%%%%%%%%%%%%%
%%%%%%%%%%%%%%%%%%%%%%%%%%%%%%%%%%%%%%%%%%%%%%%%%

%Choosing the most appropriate temporal resolution to aggregate and analyze data is a typical issue in many fields such as signal processing~\cite{SP1,SP2}, discretization of continuous variables~\cite{CV}, time series discretization~\cite{TS1,TS2}, model granularity~\cite{MG}.
%There have already been some works aiming at determining appropriate time scales for analysis of dynamic networks. The Fourier-analysis-based method presented in~\cite{Realitymining} aims at unveiling the main modes of periodicity of a dynamic network, for example circadian and weekly rhythms in human activities, which is question clearly different from the one we address here.

%\noindent\emph{Related works.}

It is paradoxical to note that, while the question of the influence of aggregation on the properties of the formed graph series is largely ignored in most of the studies on dynamic networks, this question actually already received a lot of specific attention~\cite{KKB+12,RPB13,SBK12,SBG10,CE07,STK+16,FC15,CBG11}.

%In~\cite{KKB+12}, the authors show that significant characteristics of the dynamics of a phone-call network appear at different time scales of analysis, which implies that one should use the broad spectrum of possible scales in order to observe these properties. It is very interesting to note that they do so using only statistics that are meaningful regardless of the loss of information due to aggregation, which is precisely what we are interested in here. It should be clear that the aggregation period $\gamma$ we determine is not intended to reveal the key properties of the dynamics%, it is simply the maximum aggregation period that does not alter the original dynamic network
%. Making statistics at other time scales may reveal interesting facts that are invisible otherwise. Nevertheless, for aggregation scales greater than $\gamma$, one should consider only statistics that are not sensitive to the loss of information induced by aggregation, excluding all statistics based on propagation properties of the dynamics.

In~\cite{KKB+12}, the authors lead a systematic analysis of what is visible from the structure of a dynamic phone-call network when it is aggregated at different time scales. They show that significant characteristics of the dynamics of the network appear at different scales of analysis, which implies that one should use the broad spectrum of possible scales in order to reveal these different properties of the dynamics. Though we are also concerned by the impact of the aggregation period on the properties of the formed graph series, our motivation and goal are clearly different from those of~\cite{KKB+12}. Here, we do not intend to find time scales that are able to reveal the key properties of the dynamic network or that are relevant to study phenomena taking place over the network. Instead, we aim at determining the range of aggregation scales such that the formed graph series faithfully describe the original network. Making statistics on the network out of this range of scale may still reveal interesting facts that are invisible at other scales. Nevertheless, for such aggregation scales greater than $\gamma$, one should consider only statistics that are not sensitive to the loss of information induced by aggregation (like those used in~\cite{KKB+12} for example), excluding all statistics based on propagation properties of the dynamics.

This is also the point of view developed in \cite{RPB13}, where the authors study the impact of aggregation over the properties of random walks in a dynamic network. They show that the probability of occupation of nodes of the network by such random walks is deeply impacted by aggregation, implying that it should be used with great caution when dealing with phenomenon that depends on propagation properties of the dynamics. The key contribution of the work of \cite{RPB13} is to emphasize and analytically explain the impact of aggregation on random walks, but it does not provide any way of determining a maximum aggregation period that can be used safely, which is precisely our goal here.

In \cite{SBK12}, the authors study the impact of the length of the aggregation window, as well as the impact of the type of windows used (disjoint or overlapping or starting at the beginning of the period of study), on the output of a dynamic community tracking algorithm taking as input a series of graphs. Their results show that both the length and the type of the windows used have a strong impact on the dynamic communities outputted by the algorithm. As \cite{RPB13}, the purpose of their work is to provide a deeper understanding of the effect of aggregation, but it is not intended to choose a suitable aggregation period.

Contrastingly, the goal of~\cite{SBG10} is precisely to determine an ideal aggregation period. In their method, this period is obtained as a trade-off between two metrics that vary monotonically and oppositely with regard to aggregation: one describing the loss of information (increasing with aggregation) and one describing the noise contained in the series of snapshots (decreasing with aggregation). Compared to them, here, we are concerned only with the loss of information. This allows us to avoid some drawbacks and limitations inherent to the approaches based on a trade-off: i) the value selected for the aggregation period strongly depends on the importance given to each metrics and ii) the selected value does not reveal any particular behavior of the properties of the network used in the trade-off, as each of them varies smoothly and monotonically from one extremal value to another one. On the contrary, our method does not depend on any arbitrary choice of ponderation and reveals a natural change in the way the network responds to aggregation at a certain aggregation scale that we determine.

%Secondly, it also raises questions about the metrics used for the trade-off. For example, the metrics for the loss of information depends only on the collection of snapshots and not on their order of occurrence in the series. It is therefore intrinsically unable to capture the loss of information on temporal causality of links, which is essential when aggregating the series.

\cite{CE07} also aims at determining an appropriate time scale for aggregating a link stream into a graph series. Their method does not take into account the loss of information but is instead based on the modes of periodicity and on the self-similarity of the time series of some properties of the snapshots. They observe that the offset time for which the self similarity of these time series is zero is close to half of the period of the highest frequency visible in their spectra, which is the aggregation period suggested as a result of their method. Though this provides a very relevant time scale for analyzing dynamic networks, its meaning is different from the meaning of the saturation scale we are looking for in this paper. Indeed, an important part of the activity of dynamic networks takes place at time scales much smaller than their modes of periodicity. Therefore, using such periods for aggregation usually induces an important loss of information, which we aim at avoiding here. Let us mention that a similar approach based on modes of periodicity of some time series associated to the network was previously used in~\cite{Eagle,EP06}.

The approaches of \cite{STK+16,darst2016detection} are noticeable in that they develop a method to aggregate a link stream on variable length windows. To this purpose, they fix the beginning time of the current aggregation window and they observe the evolution of some statistics of events of the aggregated graph, like the density \cite{STK+16} or the recurence of ties \cite{darst2016detection}, as the ending time of the window increases. When the observed statistics is optimized, %has converged
they end the current window and start a new one. The idea is to form a series of so-called "mature" graphs, meaning that these graphs have been aggregated on a time window long enough so that the properties of the formed graph would not change much if it was aggregated on a longer period of time.
This motivation is clearly different from ours and the loss of information due to aggregation may occur before the convergence of the properties of the formed graph.

\cite{FC15} and~\cite{CBG11} consider the aggregation of a particular class of link streams: those that are obtained as the result of the oversampling of a dynamic network where links do not occur punctually but instead last over a time interval. Such dynamic networks are often measured by sampling processes that repeatedly check (often periodically) what are the links existing in the network at different times along the period of study, e.g. using sensor devices to measure contacts between individuals \cite{EP06,CV+10}. These sampling processes introduce some noise in the data, for example due to failure to measure some links that do exist. The aim of \cite{FC15,CBG11} is to find an aggregation period that removes the noise introduced by the sampling process and allows to retrieve the original signal. Here, both our purpose and the kind of dynamic networks we consider are different. We deal only with link streams where links are punctual and do not last over time. The approach of \cite{FC15,CBG11} is not intended and not directly applicable to this kind of link streams. Conversely, it must be clear that applying our method to lasting links would require some adaptation and is one key perspective of our work. 

For sake of completeness, let us mention two other works that address in different ways the problem of aggregation of link streams into graph series. \cite{BM06} design a tool for visualizing a dynamic network as a series of snapshots, which takes as parameter the length of the aggregation window. One of the interest of this tool is that it helps to visually choose an aggregation period that gives a comprehensive view of the evolution of the network. Finally, \cite{Moo08} provides alternatives to aggregation by designing two representations of a dynamic network that encode both time and links in the form of a static graph structure.

%%%%%%%%%%%%%%%%%%%%%%%%%%%%%%%%%%%%%%%%%%%%%%%%%
%%%%%%%%%%%%%%%%%%%%%%%%%%%%%%%%%%%%%%%%%%%%%%%%%
\section{Preliminaries}\label{sec:prel}
%%%%%%%%%%%%%%%%%%%%%%%%%%%%%%%%%%%%%%%%%%%%%%%%%
%%%%%%%%%%%%%%%%%%%%%%%%%%%%%%%%%%%%%%%%%%%%%%%%%

%\marginpar{introduire vocab link stream mieux}

\begin{figure}
\begin{center}
\resizebox{.75\linewidth}{!}{
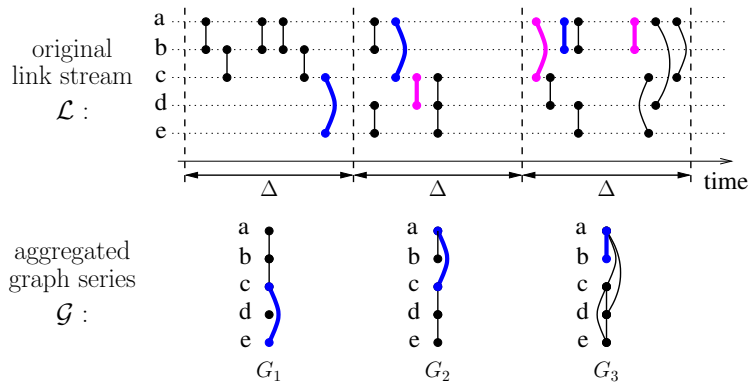
}
\end{center}
\caption{A link stream $\cur{L}$ and the graph series $\cur{G}=(G_1,G_2,G_3)$ obtained by aggregating $\cur{L}$ using a period $\Delta$. The bold dark-blue links depict a temporal path, from $e$ to $b$, in the link stream and its corresponding temporal path in the graph series. The bold light-pink links also form a temporal path in the link stream, from $d$ to $b$, but there is no temporal path from $d$ to $b$ in the graph series, because it would require to use two links of graph $G_3$, which is not allowed (see Remark~\ref{rem:strict}).\label{fig:ex-agreg}}
\end{figure}

We describe our methodology in discrete time and with non-directed links, but actually, it applies the same if the time $t$ is continuous and if the links are directed (as in the real-world datasets we consider in Section~\ref{sec:general}). The only restriction which is meaningful here is that links are punctual events and therefore have no duration. The case where links exist during one interval of time instead of one instant requires some adaptation, both for continuous and discrete time. We now formally define some of the concepts we use in the paper, starting with the process of aggregation of a link stream into a graph series, see example in Figure~\ref{fig:ex-agreg}.

\begin{definition}[Aggregation]
The aggregation, on disjoint time windows of equal length, of a link stream $\cur{L}$ on the period of study $[0,T]$ consists in choosing a constant time period $\Delta$ such that $\Delta=T/K$ for some integer $K\geq 1$ and forming the graph series $\cur{G}_{\Delta}=(G_k)_{1\leq k\leq K}$ defined by $G_k=(V,E_k)$ with $$E_k=\set{uv\ |\ \exists (u,v,t)\in \cur{L}, (k-1)\Delta\leq t< k\Delta}$$ where $V$ is the set of nodes involved in the link stream $\cur{L}$. 
%$V$ is the same for all graphs of the aggregated series: it is the set of nodes involved in the link stream $\cur{L}$. 
\end{definition}

Note that with this definition, the set of nodes $V$ of each graph of the series $\cur{G}_{\Delta}$ is the same. This is a convention adopted for convenience of description. But our methodology applies the same if one keeps, in each snapshot, only the set of nodes that have at least one link during the time window used for aggregation. Doing so, one obtains a series where the set of nodes of the graphs in the series is not fixed but depends on the time. This kind of graph series may suit better in some contexts of study. The methodological tools developped here, though described for series of graphs on a fixed set of nodes, apply also on series of graphs with a variable set of nodes.

%%%%%%%%%%%%%%%%
%%%%%%%%%%%%%%%%

A temporal path, in a link stream or a graph series, is a sequence of edges defining a path and occurring at strictly increasing time along the path (see examples given in Figure~\ref{fig:ex-agreg}).

\begin{definition}[Temporal path in a link stream]\label{def:temp-path-link-stream}
In a link stream $\cur{L}$, a temporal path $P$ is a sequence $(u_i,v_i,t_i)$ of triplets, with $1\leq i\leq l$ and $l>0$, such that $\forall i, (u_i,v_i,t_i)\in\cur{L}$ and $\forall i>1, u_i=v_{i-1}$ and $\forall i,j$, if $i<j$ then $t_i<t_j$.
\end{definition}

\begin{definition}[Temporal path in a series of graphs]\label{def:temp-path-graph-series}
In a series of graphs $\cur{G}=(G_k)_{1\leq k\leq K}$, a temporal path $P$ is a sequence $(u_i,v_i,k_i)$ of triplets, with $1\leq i\leq l$ and $l>0$, such that $\forall i, u_iv_i\in E(G_{k_i})$ and $\forall i>1, u_i=v_{i-1}$ and $\forall i,j$, if $i<j$ then $k_i<k_j$.
\end{definition}

\begin{remark}\label{rem:strict}
Note that, in Definitions~\ref{def:temp-path-link-stream} and~\ref{def:temp-path-graph-series}, the inequalities are strict. This implies that a temporal path cannot use two links belonging to the same graph of the series or occurring at the same time in the link stream.
\end{remark}

Temporal paths are an essential notion as they capture the propagation properties of the dynamic network. Indeed, all diffusion phenomena in the network, such as communication of information, spreading of epidemics and cascades of influence for example, respect time causality: a node needs to be reached by the diffusion before it can propagate it further. Therefore, all these phenomena occur on and follow temporal paths of the dynamic network.

There are two notions of length associated to a temporal path: the topological length, which is the classical one for static graphs, and the duration of the path.

\begin{definition}[hops(P) and time(P)]
In a link stream or a graph series, the \emph{topological length} of a temporal path $P=((u_i,v_i,t_i))_{1\leq i\leq l}$ is the number $l$ of edges in the path. In the rest of the paper, we call it the \emph{number of hops} of $P$ and denote it $hops(P)$.\\
The \emph{duration} of path $P$, denoted $time(P)$, is $t_l-t_1$ in a link stream and $t_l-t_1+1$ in a graph series (because each $t_i$ is not an instant as in a link stream but an interval of time which has a duration).
\end{definition}

%It is worth to note the following.

\begin{remark}\label{rem:hops}
By definition, in a graph series, we always have $hops(P)\leq time(P)$ for any temporal path $P$. Note that this does not hold for link streams in general, since time is not necessarily indexed by an integer.
\end{remark}

In the rest of the paper, we also use three notions of distance at time $t$ between two nodes $u,v$ of a link stream or a graph series. These notions are based on the minimal arrival time $t_{arr}$, if any (otherwise $t_{arr}$ is undefined), among all paths from $u$ to $v$ whose departure time is not before $t$.
The \emph{distance in time}, denoted $d_{time}(u,v,t)$, is simply defined as $t_{arr}-t$ in a link stream and $t_{arr}-t+1$ in a graph series, with the convention $d_{time}(u,v,t)=+\infty$ if $t_{arr}$ is undefined. The \emph{distance in hops}, denoted $d_{hops}(u,v,t)$, is the minimum number of hops among all paths realizing the distance in time $d_{time}(u,v,t)$. By convention, $d_{hops}(u,v,t)=+\infty$ when $d_{time}(u,v,t)=+\infty$.
Finally, the distance in absolute time, which is dedicated to aggregated graph series only, is denoted $d^{abs}_{time}(u,v,t)$ and defined by $d^{abs}_{time}(u,v,t)=\Delta . d_{time}(u,v,t)$. It is the absolute time needed to go from node $u$ to node $v$ in the aggregated graph series, with a departure time not before $t$, taking into account the fact that each graph of the series represent a time interval of length $\Delta$. Interestingly, this last notion contains in itself the imprecision of the timestamps of the aggregated series.

%%%%%%%%%%%%%%%%%%%%%%%%%%%%%%%%%%%%%%%%%%%%%%%%%
%%%%%%%%%%%%%%%%%%%%%%%%%%%%%%%%%%%%%%%%%%%%%%%%%
\section{Difficulty of the problem} \label{sec:difficult}
%%%%%%%%%%%%%%%%%%%%%%%%%%%%%%%%%%%%%%%%%%%%%%%%%
%%%%%%%%%%%%%%%%%%%%%%%%%%%%%%%%%%%%%%%%%%%%%%%%%

%\marginpar{est-ce que ça a ete assez bien explique avant?}

\begin{figure}
\begin{center}
   \includegraphics[width=0.49\textwidth]{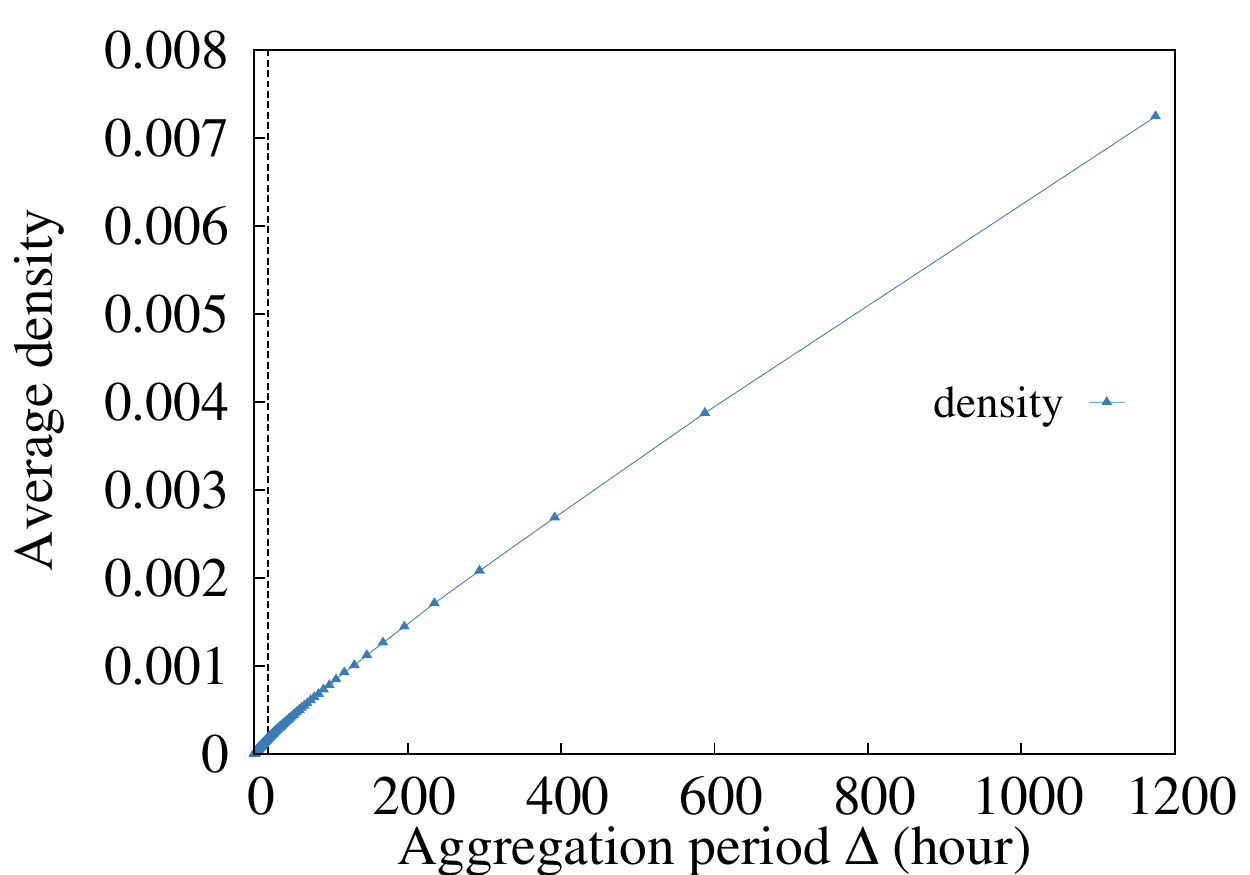}
   \includegraphics[width=0.49\linewidth]{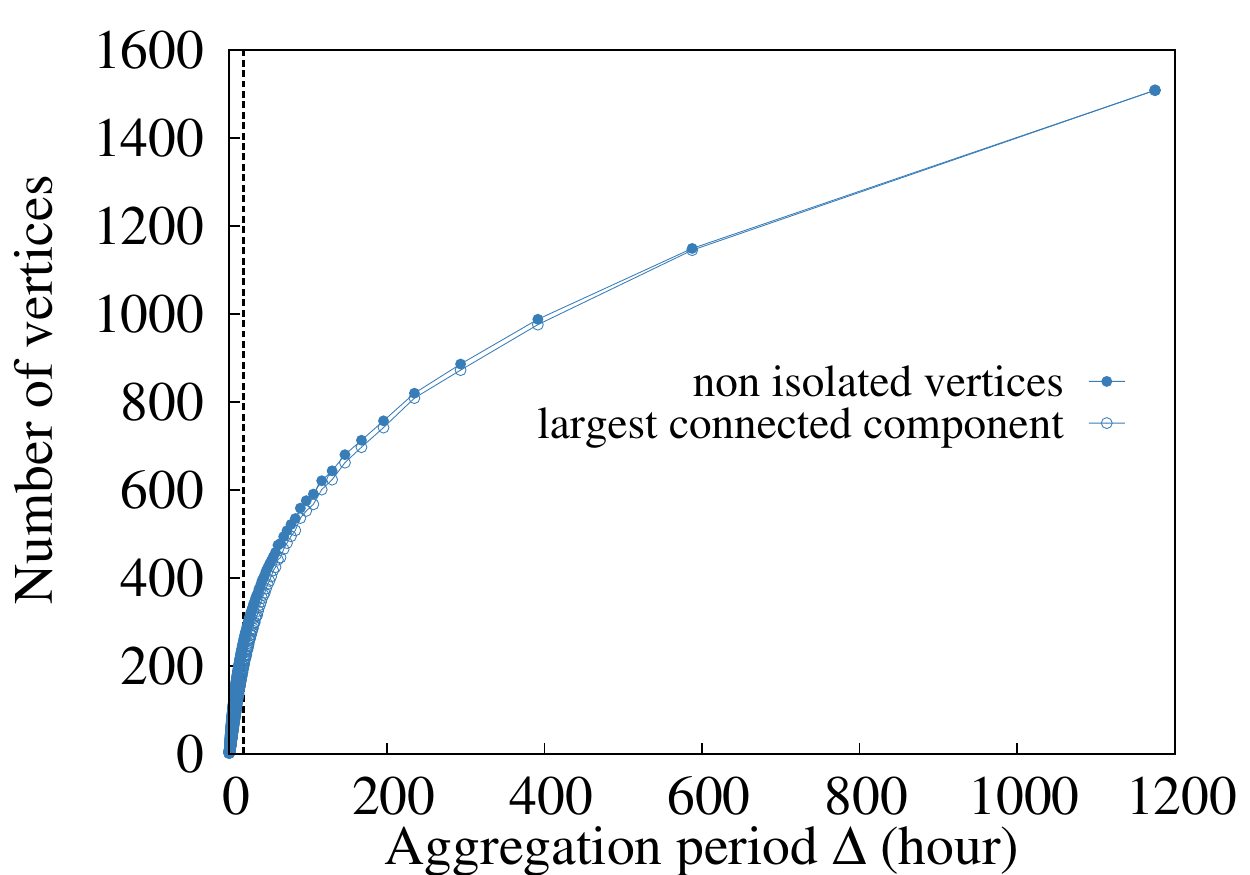}
   \includegraphics[width=0.49\linewidth]{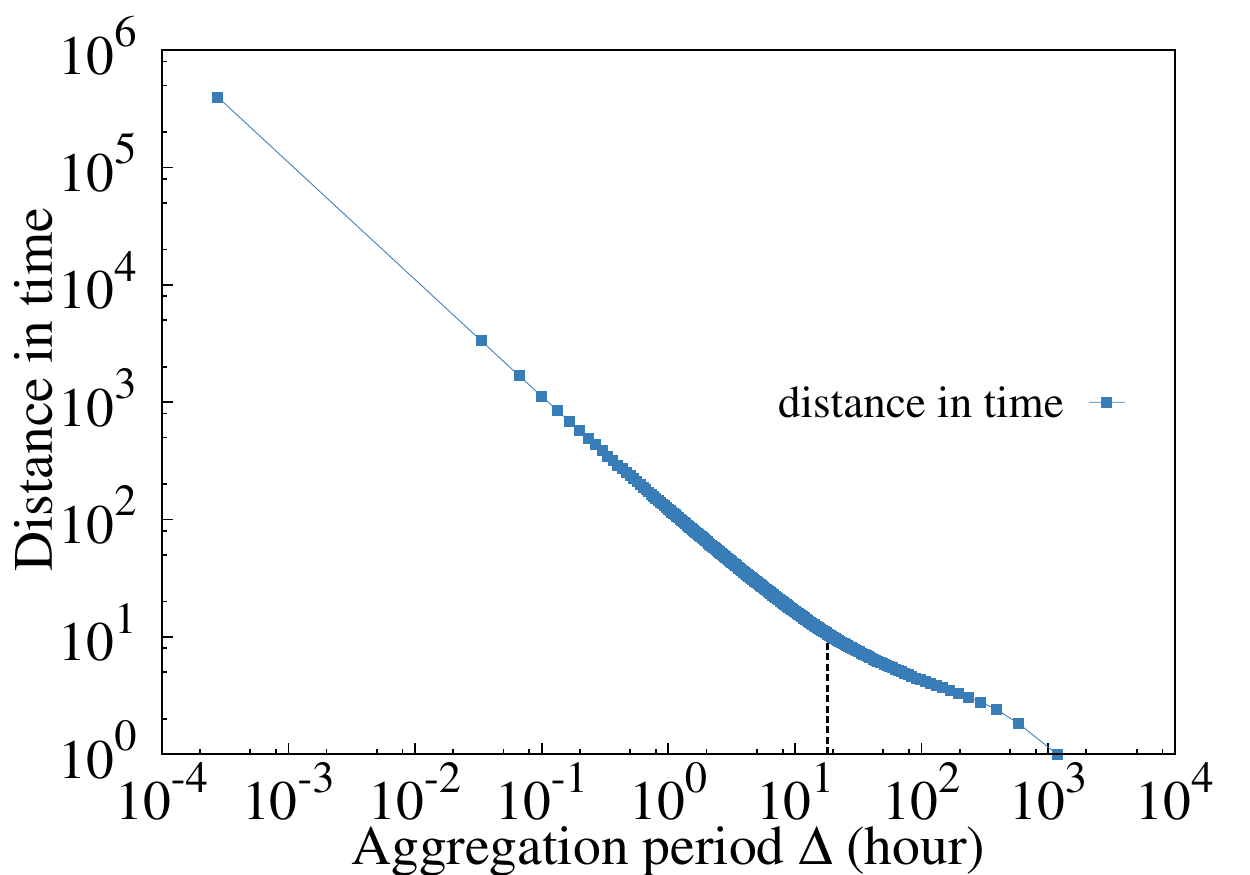}
   \includegraphics[width=0.49\linewidth]{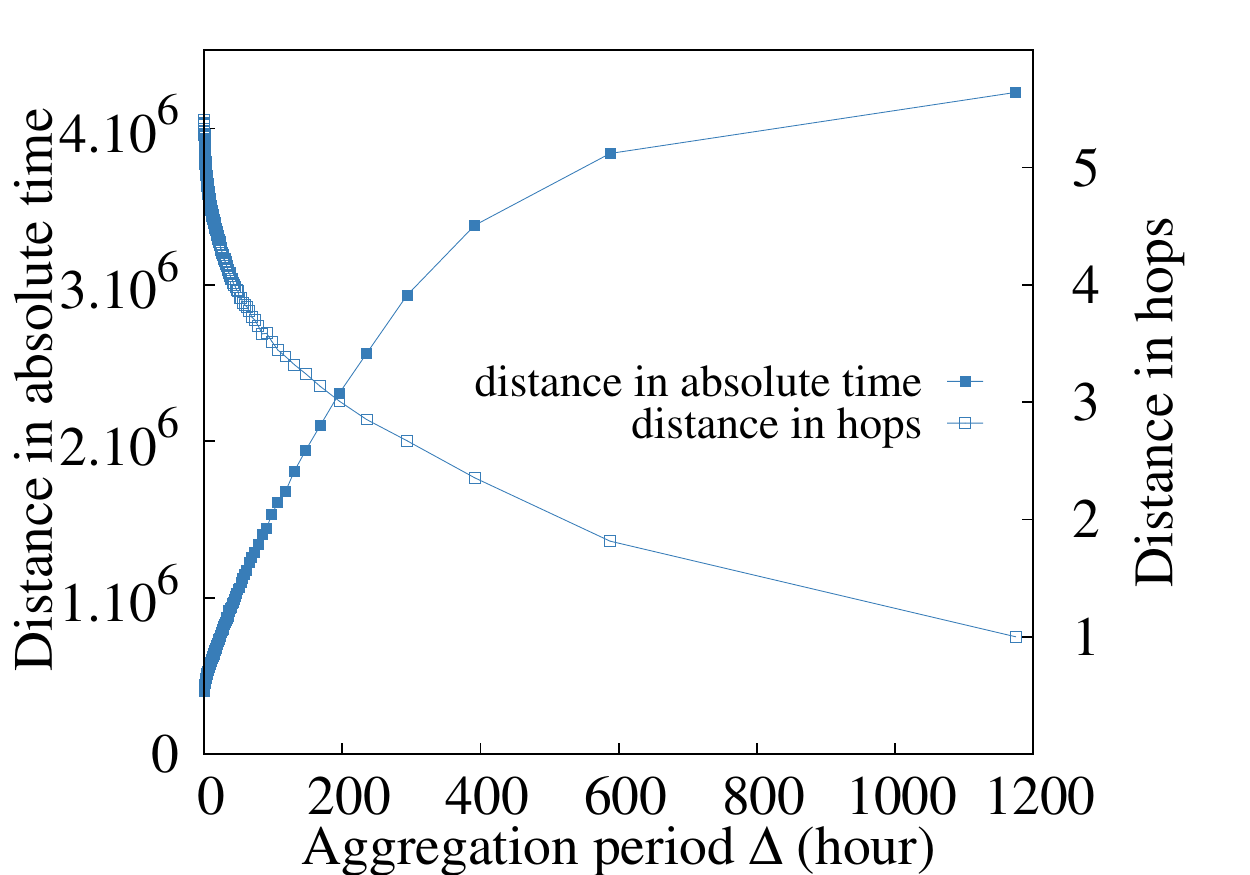}
\caption{Variation of some classical parameters of the aggregated series of graphs (y-axis) according to the aggregation period $\Delta$ (x-axis), for the Irvine network (see Section~\ref{sec:general}). 
%All curves are normalized to have maximum value $1$.
Top-left: density. Top-right: connectedness properties. Bottom-left: distance in time (log-log scale). Bottom-right: other distance properties (linear scale). The dotted line shows the aggregation period returned by the occupancy method: 18h.}
\label{fig:param}
\end{center}
\end{figure}

As previously explained, the larger the length of the aggregation window, the greater the loss of temporal information due to aggregation, as the exact times of occurrence of the links within one given window are lost. Then, we can reformulate the problem as follows: what is the maximum aggregation period that induces no significant loss of information in the graph series compared to the original link stream? % Or conversely, we want to determine what is the minimum aggregation period $\gamma_{alt}$ such that the formed graph series shows evidence of alteration.
A natural way to proceed to answer this question is to make the aggregation period vary from its minimal value to its maximal value and to observe the variations of the properties of the obtained series of graphs in the meanwhile. Then, one would hope to find a time scale beyond which the variation of these properties exhibit a qualitative change. Unfortunately, this does not happen for the classical properties of interest of the graph series. On the opposite, when the aggregation period varies, these properties varies smoothly from one extremal value to another one. Figure~\ref{fig:param} shows the results for several properties of the Irvine network, which we use as an example to describe our method in the first sections of this article (see Section~\ref{sec:general} for a description of the Irvine dataset).

%\marginpar{ou est la distance en temps normale?}
%\marginpar{quel rational pour la distance temps corrigee?}
%\marginpar{denormaliser les courbes?}
%\marginpar{mettre la dist time en log log?}

%%%%%%%%%%%%%%%%
%%%%%%%%%%%%%%%%

Figure~\ref{fig:param} top-left shows the variations of the mean density of the snapshots of the series, which is also equivalent, up to a multiplicative factor $n-1$ (where $n$ is the number of vertices), to the mean degree of the nodes in all the snapshots. The plot shows that when the aggregation goes from the minimal temporal resolution of the timestamps (1s) to the total length of the period of study ($\sim$1175h), these two properties linearly varies from a very small value ($5.7\times 10^{-7}$) to their maximal value ($7.2\times 10^{-3}$), which is the one obtained by aggregating the whole dynamic network into one single graph. In this case, the mean density of the graphs in the series is therefore equal to the density of the totally aggregated graph.

The plot in Figure~\ref{fig:param} top-right shows that in the meanwhile the mean size of the largest connected component in each snapshot as well as the mean number of non isolated vertices per snapshot (which are very close) exhibit the same behavior: their values go increasingly from a minimal one ($2.3$ nodes) to a maximal one ($1509$ nodes), which is the total number of nodes in the network, without exhibiting any non-smooth behavior at any time scale.

Let us now examine the variations of distance properties according to the aggregation period. Figure~\ref{fig:param} bottom-left gives the variation of the mean distance in time $d_{time}(u,v,t)$ (see Section~\ref{sec:prel}) for all couples $(u,v)$ of nodes and all time $t$ (such that $d_{time}(u,v,t)$ is finite), in logarithmic scale. As one can see, the curve is almost a straight line, indicating that the mean distance in time depends on $\Delta$ following a power law. This comes from the fact that the number of graphs formed in the aggregated series varies as $1/\Delta$: the plot shows that the mean distance in time varies accordingly. This does not help much to detect a time scale at which the properties of the aggregated series significantly change their behavior.

Pushing further, in Figure~\ref{fig:param} bottom-right, we also plotted the mean distance in hops (empty squares) and the mean distance in absolute time (filled squares), both in linear scale. The rational for using the distance in absolute time, defined as $d^{abs}_{time}(u,v,t)=\Delta . d_{time}(u,v,t)$, is that it does not suffer from the dependence on $1/\Delta$ previously highlighted for the distance in time, since it is canceled by the multiplication by $\Delta$. Then, it gives a clearer insight into the variations of the distance in time with the aggregation period. Unfortunately, as one can see, the situation is the same as for the other parameters previously studied. When the aggregation period $\Delta$ increases from the minimal temporal resolution of the timestamps (1s) to the total length of the period of study ($\sim$1175h), the mean distance in absolute time increases as well, going monotonically from its minimal value ($\sim$110h) until its maximal value ($\sim$1175h), which is by definition equal to the total length of the period of study, as there is only one graph in the series formed using the maximum value of $\Delta$. In the meanwhile the mean distance in hops (empty squares in Figure~\ref{fig:param} bottom-right) decreases, from $5.4$ to $1$, without exhibiting any remarkable change at any value of the aggregation period.

Thus, the observation of the variations of the classical properties of the graph series with the aggregation period does not point out scale at which some qualitative changes occur in the way the dynamic network responds to aggregation. Instead, one finds a regular drift from one extreme value to another one\footnote{Note that we present results for only one dataset but they hold similarly for all the four datasets we consider in this paper, cf. Section~\ref{sec:general}.}. This constitutes the main difficulty of the problem we consider. 

%\textcolor{red}{The temporal information can be seen as the order of successive events. Considering two temporal contacts $(u,v,t_1)$ and $(v,w,t_2)$ with $t_1<t_2$, one notes that the path $((u,v,t_1), (v,w,t_2))$ exists. Yet, if after the aggregation, the two links belong to the same time window, one will lose the information about their time ordering. The information about the existence of such a path is not anymore accessible. This simple example let understand how much the temporal information plays a crucial role during a spreading process. Therefore, in order to catch the link between the aggregation time and the temporal loss, it seems inappropriate to study static parameters for each graph. An important contribution of our work is to exhibit a finer  property of the graph series based on temporal paths that is able to reveal a time scale where such qualitative changes occur.}

%%%%%%%%%%%%%%%%
\begin{figure}
\begin{center}
   \includegraphics[width=0.49\textwidth]{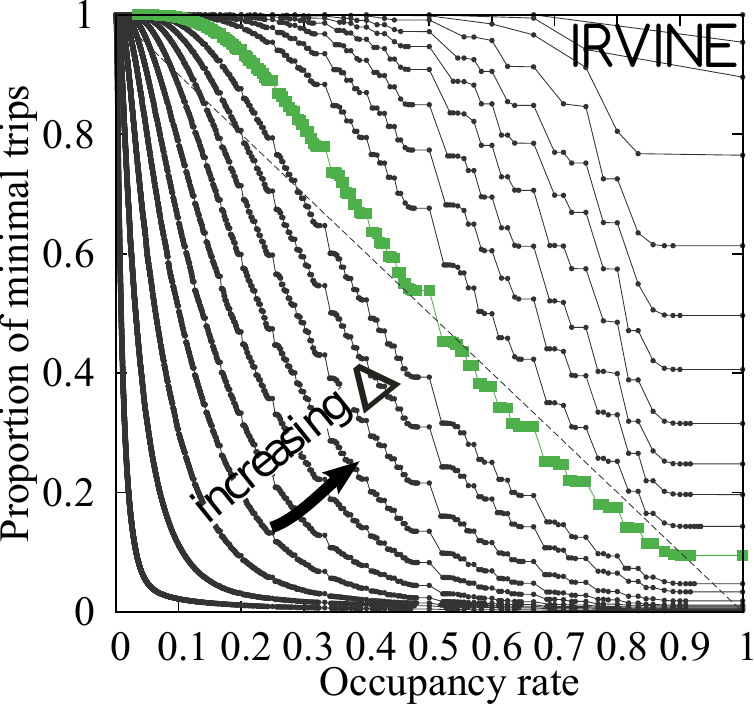}
   \includegraphics[width=0.49\textwidth]{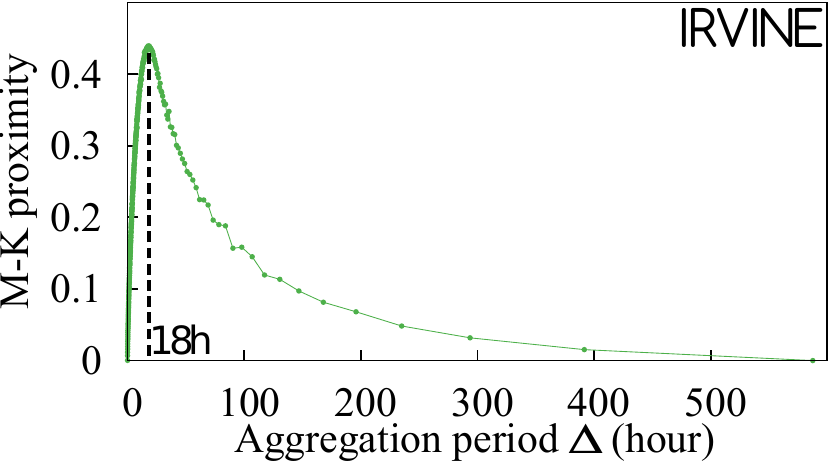}
\caption{Left: Inverse Cumulative Distributions (ICD) of the occupancy rates (x-axis) of the minimal trips of the aggregated series $\cur{G}_{\Delta}$ for several values of the aggregation period $\Delta$ in the range $\inter{1,T}$, for the Irvine network. Right: M-K proximity (y-axis) of these distributions with the uniform density distribution according to $\Delta$ (x-axis).
}
\label{fig:method}\label{fig:method-max}
\end{center}
\end{figure}
%%%%%%%%%%%%%%%%

%%%%%%%%%%%%%%%%%%%%%%%%%%%%%%%%%%%%%%%%%%%%%%%%%
%%%%%%%%%%%%%%%%%%%%%%%%%%%%%%%%%%%%%%%%%%%%%%%%%
\section{The occupancy method} \label{sec:method}
%%%%%%%%%%%%%%%%%%%%%%%%%%%%%%%%%%%%%%%%%%%%%%%%%
%%%%%%%%%%%%%%%%%%%%%%%%%%%%%%%%%%%%%%%%%%%%%%%%%

%Our aim is to determine the maximum aggregation period that induces no significative distorsion of the properties of the link stream. The natural way to proceed to do so is to make the aggregation period vary from its minimal value to its maximal one and to observe how the properties of the obtained series of graphs varies in the meanwhile. Then, one would expect to find a scale of time beyond which these properties exhibit a qualitative change in the way they vary, indicating a change in a way the network respond to aggregation. Unfortunately, as shown on Fig.~\ref{fig:param}, the classical properties one may have in mind, such as density, degrees, connectedness, distances in time and in hops, do not exhibit such a behavior: they vary smoothly from one extremal value to another one.

%\begin{sloppypar}
We now give the definitions necessary to describe our method and we illustrate it on a sample real-world network, the Irvine network (cf. Section~\ref{sec:general}).
%\end{sloppypar}

%In this part, we exhibit a parameter of the dynamic network that, unlike the classical parameters studied above, does not monotonically varies when the aggregation period increases. Instead, this parameter starts from a low value, then grows until a maximum value, reached for a value of the aggregation period denoted $\gamma$, and finally decreases until reaching again a very low value. This behavior distinguishes two ranges of aggregation period: the range below $\gamma$, where the properties of propagation of the link streams are mainly conserved in the aggregated series of graphs, and the range beyond $\gamma$, where these properties show evidence of alteration.

\begin{definition}[Trip and minimal trip]
A \emph{trip} is a quadruplet $(u,v,t_{dep},t_{arr})$ such that there exists a temporal path from $u$ to $v$ whose starting time from $u$ and arriving time at $v$ are both in the interval $[t_{dep},t_{arr}]$. A trip $(u,v,t_{dep},t_{arr})$ is \emph{minimal} if there exists no trip from $u$ to $v$ in an interval $[t'_{dep},t'_{arr}]$ strictly included in $[t_{dep},t_{arr}]$ (i.e. $[t'_{dep},t'_{arr}]\subsetneq [t_{dep},t_{arr}]$).
\end{definition}

\begin{definition}[Transition and shortest transition]\label{def:trans}
A temporal path $P$ on two hops, i.e. $P=((a,b,t_1),(b,c,t_2))$, is called a \emph{transition}, and $P$ is a \emph{shortest transition} if $(a,c,t_1,t_2)$ is a minimal trip.
\end{definition}

\begin{definition}[Occupancy rate]
%\begin{sloppypar}
For a graph series $\cur{G}$ and a temporal path $P$ in $\cur{G}$, the \emph{occupancy rate} of path $P$, denoted $occ(P)$, is defined as $occ(P)=hops(P)/time(P)$.
The occupancy rate of a minimal trip $(u,v,$ $t_{dep},t_{arr})$ is the occupancy rate of a temporal path starting from $u$ at $t_{dep}$ and arriving at $v$ at $t_{arr}$ and having the minimum number of hops among such paths. Note that we always have $0<occ(P)\leq 1$, since from Remark~\ref{rem:hops}, $0<hops(P)\leq time(P)$.
%\end{sloppypar}
\end{definition}

The rational behind the occupancy rate $occ(P)$, which is comprised between $0$ and $1$, is to count the proportion of time steps between $t_{dep}$ and $t_{arr}$ that are effectively used by path $P$ to move from one node of the dynamic network to another one. Indeed, only some of the graphs $G_t$ with $t_{dep}\leq t\leq t_{arr}$ contain a link of path $P$, but not all of them. Therefore, a path $P$ uses some time steps to move to the next node on the path and spends the rest of the time steps simply waiting on the node reached so far, until the next hop to be performed on path $P$ occurs. Then, in other words, the occupancy rate quantifies how much the path $P$ is busy moving from one node to the next one, taking into account that $P$ also spends some proportion of the time without moving. As we explain later in this section, the occupancy rate of paths in the agregated graph series is strongly impacted by the aggregation period used to form the series, and it is at the core of our method to determine the saturation scale $\gamma$.

%\textcolor{red}{The occupancy rate parameter can also be defined as the frequency of the average time of the successive transitions that constitute the path $occ(P)=\frac{l}{(t_2-t_1+1)+..+(t_l-t_{l-1}+1)}$. Therefore, the occupancy rate parameter is directly linked to the transitions and so the temporal information of the graph series.} 
% Intuively, it can be seen as the compression ratio of the temporal information, it provides some insights about the local alteration. If the 

To this purpose, we make the aggregation period $\Delta$ vary from its minimal value, the resolution of the timestamps, until the whole length $T$ of study of the network. For each value of $\Delta$ we form the aggregated graph series $\cur{G}_{\Delta}$ for which we compute the set of minimal trips and their occupancy rates. Then, for each $\Delta$, we plot the distribution of occupancy rates of all the minimal trips in $\cur{G}_{\Delta}$ (considering all pairs of nodes and all time intervals), see Figure~\ref{fig:method} left.

%We also plot the M-K distance (y-axis) between each of these distributions and the uniform density distribution on $[0,1]$ as a function of $\Delta$ (x-axis), see Figure~\ref{fig:method} right.
 
%We consider only the minimal trips $(u,v,t_{dep},t_{arr})$ such that $u$ and $v$ are not in contact on time interval $[t_{dep},t_{arr}]$.
%\marginpar{QY : by the way, ils y sont les single link dans les faisceaux de distribs?}

%%%%%%%%%%%%%%%%
\begin{figure}
\begin{center}
%\null
%\hfill
     %\includegraphics[width=0.24\linewidth]{./irvine_faisceaux2.pdf}
     \includegraphics[width=0.32\linewidth]{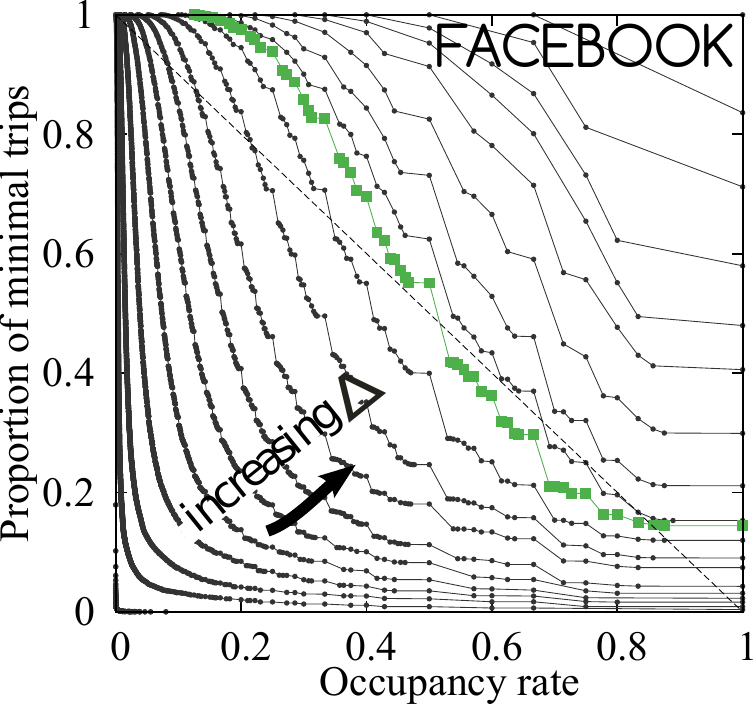}
%\hfill
     \includegraphics[width=0.32\linewidth]{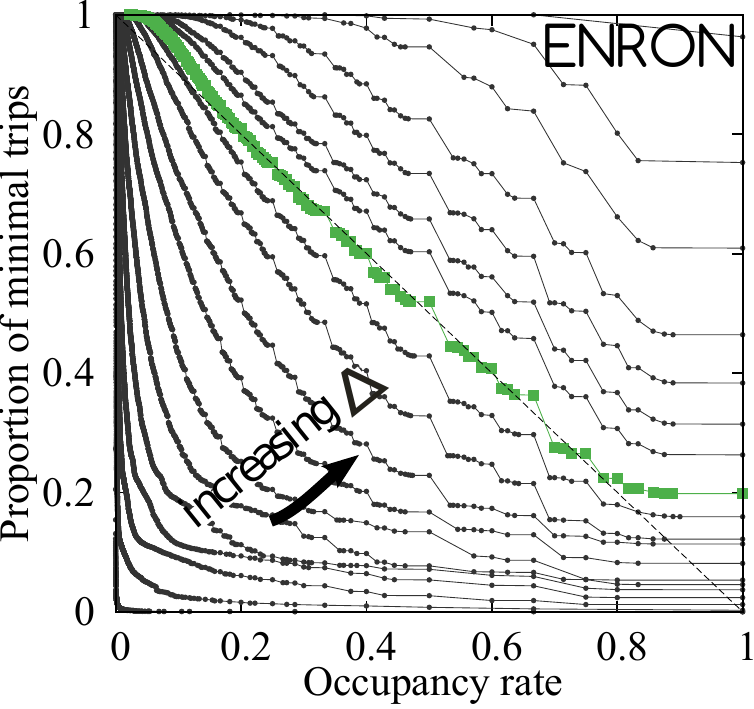}
%\hfill
     \includegraphics[width=0.32\linewidth]{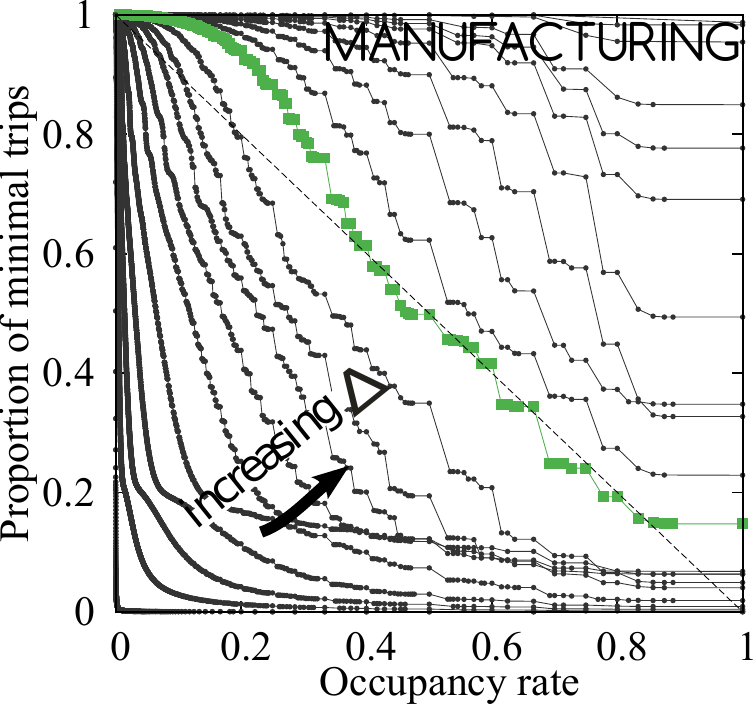}
%\hfill
%\null
%\label{fig:method}
\end{center}
\caption{Inverse Cumulative Distributions (ICD) of the occupancy rates (x-axis) of the minimal trips of the aggregated series $\cur{G}_{\Delta}$ for several values of the aggregation period $\Delta$ in the range $\inter{1,T}$, for the Facebook, Enron and Manufacturing networks.}
\label{fig:icd_mk}
\end{figure}
%%%%%%%%%%%%%%%%

Necessarily, when $\Delta$ is close to its minimal value, provided that the resolution of the timestamps is fine enough, the distribution of occupancy rates must be concentrated on values close to $0$. The reason is that the aggregation windows contain only few data and the shortest paths therefore need to wait several slot of times before finding one opportunity to perform the next hop. On the opposite, when the aggregation period reaches its maximum value, by definition, all the minimal trips are made of one single link (because there is only one graph in the aggregated series) and their occupation rate is $1$. Then, the distribution is again concentrated, this time on the value $1$. What is remarkable here (Figure~\ref{fig:method} left) is that the distribution changes from values concentrated near $0$ to values concentrated on $1$ in a very specific manner: it first progressively stretches toward $1$ until it almost equally occupies all the values on the range from $0$ to $1$ and then it contracts again, leaving the low values to progressively concentrate on the values close to $1$.

The saturation scale $\gamma$ is precisely the value of $\Delta$ for which the distribution is maximally stretched on the interval $[0,1]$ (curve marked with green squares on Figure~\ref{fig:method} left). In order to detect it, we compute for each value of $\Delta$ in the total range of variation, the M-K distance $d(\Delta)$ (see Section~\ref{sec:selec} for a definition) between the distribution obtained for $\Delta$ and the uniform density distribution on $[0,1]$, i.e. the distribution whose inverse cumulative is the straight line $y=1-x$. We then plot the M-K proximity, defined as $1/2-d(\Delta)$, in Figure~\ref{fig:method-max} right. This confirms the observation made above on the way the distribution first stretches and then concentrate again: accordingly, the M-K proximity first increases and then decreases. As a consequence, the value $\gamma$ returned by our method is the value of $\Delta$ that realizes the maximum of the M-K proximity. Of course, one may think of many other ways to determine which $\Delta$ gives the maximum stretch of the distribution. We actually tried several of them (see Section~\ref{sec:selec}) and we chose to use the M-K distance with the uniform density distribution because it gives results that are visually satisfying and it is conceptually simple.

Let us now explain the meaning of $\gamma$. As pointed out in the introduction, the most significative loss of information due to aggregation is the loss of the order in which the links involving one given node $u$ occur in a given aggregation window. This loss makes it impossible to know whether there exists a transition from node $v_1$ to node $v_2$ going through node $u$ within an aggregation window where both links $(v_1,u)$ and $(u,v_2)$ occur: this depends on whether $(v_1,u)$ occurs before $(u,v_2)$ or not, which is lost with aggregation. Then, the propagation properties of the aggregated series do not faithfully reflect those of the original link stream.

To this regard, a very low occupancy rate for most minimal trips of the series denotes that most paths spend a long time waiting between two consecutive hops. This means that there are only very few links in each aggregation window, which implies that only very few information is lost in the whole graph series about the order of occurences of links. On the opposite, a very high occupancy rate for most of the minimal trips reveals that at each time in the graph series, there is a high probability to find a next hop to perform on any given shortest path, meaning that, in each snapshot, a high proportion of nodes are involved in a high number of edges. Then, at the same time, the information on the existence or the non existence of a transition, in the original link stream, using a couple of these edges incident to one same node is lost, which constitutes the essential loss resulting from the aggregation process.

What makes the occupancy rate so remarkable with regard to aggregation, compared to the classical parameters such as density for example (see Figure~\ref{fig:param}), is that the evolution of the distribution of occupancy rates when the aggregation period varies clearly shows two distinct phases. In the first phase of variation, below $\gamma$, only the low values of the distribution increase, while the proportion of high occupancy rates almost does not change. This means that during this phase, the effect of increasing the aggregation period is mainly to fill the lack of links in the aggregation windows without inducing a significant loss of information. On the opposite, in the second phase, beyond $\gamma$, there is a strong increase of the proportion of minimal trips having a very high occupancy rate, $1$ or close to $1$, indicating that the loss of information due to aggregation becomes non-negligible. Therefore, the saturation scale $\gamma$ appears as a separation between the range of values, below $\gamma$, where the aggregated graph series still faithfully describes the original link stream and the range of values, beyond $\gamma$, where aggregation alters the properties of propagation of the original link stream.

%%%%%%%%%%%%%%%%
\begin{figure}
\begin{center}
%\null
%\hfill
   %\includegraphics[width=0.24\linewidth]{./max_mk_irvine.pdf}
   \includegraphics[width=0.32\linewidth]{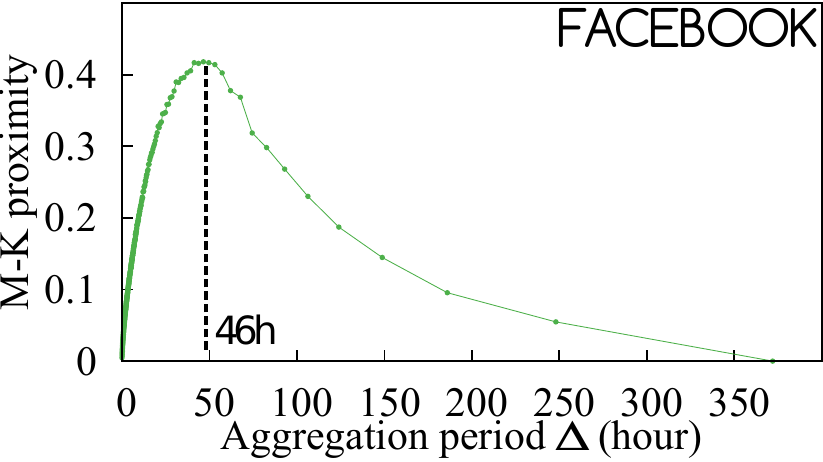}
%\hfill
   \includegraphics[width=0.32\linewidth]{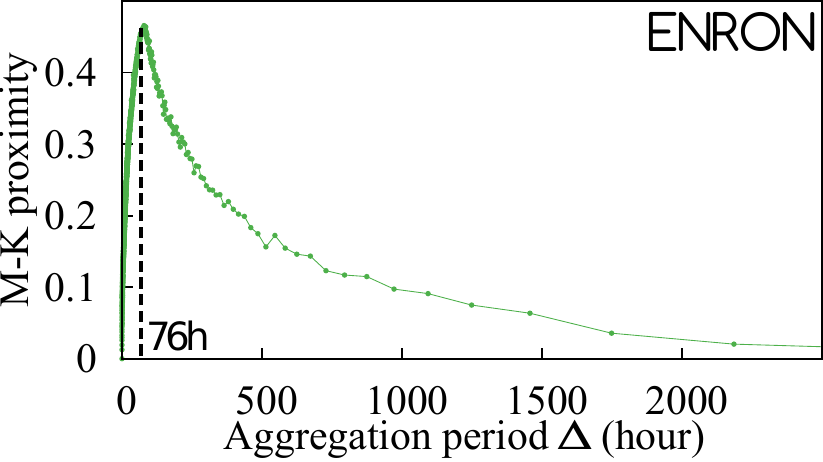}
%\hfill
   \includegraphics[width=0.32\linewidth]{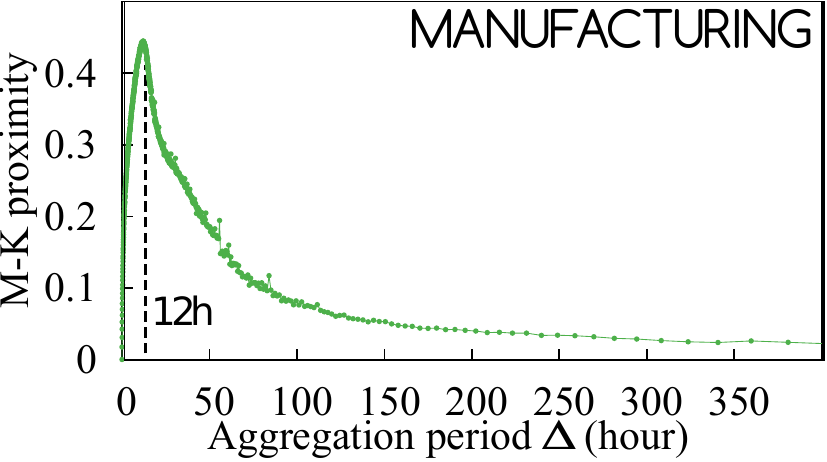}
%\hfill
%\null
\end{center}
\caption{M-K proximity (y-axis) of the distribution of occupancy rates of minimal trips of the aggregated series $\cur{G}_{\Delta}$ according to the aggregation period $\Delta$ (x-axis), for the Facebook, Enron and Manufacturing networks.}
\label{fig:max_mk}
%\label{fig:method-max}
\end{figure}
%%%%%%%%%%%%%%%%

%%%%%%%%%%%%%%%%%%%%%%%%%%%%%%%%%%%%%%%%%%%%%%%%%
%%%%%%%%%%%%%%%%%%%%%%%%%%%%%%%%%%%%%%%%%%%%%%%%%
\section{Results on real-world data\-sets} \label{sec:general}
%%%%%%%%%%%%%%%%%%%%%%%%%%%%%%%%%%%%%%%%%%%%%%%%%
%%%%%%%%%%%%%%%%%%%%%%%%%%%%%%%%%%%%%%%%%%%%%%%%%

%\marginpar{decider partout entre our method et the occupancy method}

In this section we apply our methodology and discuss the results obtained on four link streams, whose timestamps have a resolution of 1s.
The \emph{UC Irvine messages} network~\cite{Irvine}, which is the one used for presentation of the method in the previous section, is made of 48\,000 messages sent between 1\,509 users of an online community of students from the University of California, Irvine, over a period of 48 days. % 0.66 msg/day/user : 18.7h
The \emph{Facebook wall posts} network~\cite{VMC+09} is made of 11\,991 wall posts between a group of 3\,387 Facebook users over a period of 1 month. % 0.12 msg/day/user agreg: 53h
The \emph{Enron emails} network~\cite{KY04} contains 15\,951 individual emails sent between a group of 150 employees of the Enron company during year 2001. % 0.29 msg/day/user agreg: 60h
Finally, the \emph{Manufacturing emails} network~\cite{Manufacturing} contains 82\,894 internal emails between 153 employees of a mid-sized manufacturing company over a period of 8 months. % 2.22 msg/day/user agreg: 11h

From a computational point of view, let us mention that the algorithm we use for computing the distribution of occupancy rates in one given graph series $\cur{G}$ has time complexity $O(nM)$, where $n$ is the number of vertices of one graph in the series (which is the same for all graphs) and $M$ is the total number of edges in all the graphs of the series. It is a dynamic programming scheme going backward in time: at one step, knowing all the minimal trips of the series starting not before time $k+1$, the algorithm computes the minimal trips starting exactly at time $k$, their duration and their minimum number of hops. In the occupancy method, this algorithm has to be run as many times as the number of values of $\Delta$ used for the aggregation period. But the most costly computations are the ones made for small values of $\Delta$, as $M$ is then large.

We applied the occupancy method on each of the four datasets mentionned above. The distributions of occupancy rates of the minimal trips in the aggregated graph series are given on Figure~\ref{fig:method} left for Irvine and Figure~\ref{fig:icd_mk} for the three other networks, their M-K proximity with the uniform density distribution is given on Figure~\ref{fig:method-max} right and Figure~\ref{fig:max_mk}.
One can see that the observations made on the Irvine network in Section~\ref{sec:method}, % concerning the way the distribution of occupancy rates evolves with the aggregation period
hold for all the four datasets. When the aggregation period $\Delta$ increases, the distribution of occupancy rates, initially concentrated near $0$, stretches until it occupies almost equitably all the range of values between $0$ and $1$, and then concentrates again on the values close to $1$. Consequently, the proximity with the uniform density distribution first increases, until it reaches a maximum for $\Delta=\gamma$, which is the saturation scale returned by the occupancy method, and then decreases until the aggregation period reaches its maximum value $T$.
This shows that the way the distribution of occupancy rates evolves with the aggregation period is a fundamental phenomenon common to many dynamic networks, therefore guaranteeing that our method is sound and that it can be used for a wide range of dynamic networks.

The values returned for $\gamma$ in each of the four cases are: 18 hours for the Irvine message network, 46 hours for the Facebook wall-post network, 78 hours for the Enron email network and 12 hours for the Manufacturing email network. These values, between half a day and three days, are in accordance with the fact that both emails and on-line social network messages are generally not dedicated to live discussions. In the case of email networks for example, most of people only send some emails a day and frequently wait for some hours or some days before getting a reply. Therefore, this range of values seems appropriate for the largest aggregation scales providing accurate views of the original link streams.

The aggregation periods returned by our method also appear to be in accordance with the level of activity of these 4 networks. The two greater values, 46h for Facebook and 78h for Enron, are obtained for the two networks that have the lower activity, $0.12$ and $0.29$ messages sent in average per person per day for Facebook and Enron respectively. The two other networks have higher activities, $0.66$ messages per person per day in the Irvine network and $2.22$ in the Manufacturing network, and have smaller saturation scales, 18 hours and 12 hours respectively. As one can see, the average activity has a strong influence on the saturation scale, even though this is not the only parameter affecting it. We further investigate the relationship between the level of activity and the saturation scale in the next section.%Section~\ref{sec:synth}.

Finally, let us emphasize that the aggregation period $\gamma$ returned by our method should not been interpreted as the best possible one but instead as an upper-bound on the aggregation periods that are suitable for studying the network. For many practical studies, one may prefer to choose an aggregation period slightly lower than $\gamma$, which will preserve more carefully the properties of the network. For example, in the case of the four networks we study here, one can note that the proportion of minimal trips having occupancy rate $1$ started to increase just before the distribution of occupancy rates reaches its maximal stretched position (the one selected by our method). Then, one could prefer to use an aggregation period smaller than $\gamma$ in order to get a finer grain representation of the dynamic network. In Section~\ref{sec:valid}, we give some ways to directly estimate the loss of information in the aggregated graph series that can be used to choose more accurately the aggregation period in the range of scales immediately preceding $\gamma$.

%%%%%%%%%%%%%%%%%%%%%%%%%%%%%%%%%%%%%%%%%%%%%%%%%
%%%%%%%%%%%%%%%%%%%%%%%%%%%%%%%%%%%%%%%%%%%%%%%%%
\section{Results on synthetic networks}\label{sec:synth}
%%%%%%%%%%%%%%%%%%%%%%%%%%%%%%%%%%%%%%%%%%%%%%%%%
%%%%%%%%%%%%%%%%%%%%%%%%%%%%%%%%%%%%%%%%%%%%%%%%%

% Pour uniforme:
% faire les courbes avec beaucoup plus de points ~20 : discuter seulement selon l'intercontact moyen qui est exactement l'inverse de l'activité tel qu'utilisé plus haut dans le papier: nb_cont/pers/temps  
% evt : à activité fixée et à taille du réseau fixée, étudier la dépendance à la durée de l'expérience, c.à.d. en faisant varier T.
% evt : à activité fixée et à durée fixée, étudier la dépendance à la taille du réseau. Il y a 2 façon de fixer l'activitée: fixer le degré, ou fixer la densité. Essayer les 2. Voir laquelle est la bonne et la prendre pour la suite.

% Pour nuit/jour
% il y a déjà les courbes dont on a besoin. Il faut simplement:
% rajouter des points dans l'intervalle sans point
% séparer des uniformes, garder les réels?, ne pas rescaler mais savoir si on met en absisce l'activité basée sur degrés ou sur densité?
% on fait 3 courbes, avec une seule notion d'activité, on fait 1) moyenne et 2) médian (avoir si c'est utile si on rescale pas), la 3 ème courbe c'est l'agreg en fonction du pourcentage de nuit, elle y est déjà aussi
% Sur la 3ème courbe, il faut mettre aussi la moyenne et le médian desintercontacts

% pour les données réelles, on peut eventuellement montrer les non-differentiables à 2 sauts en plus, si c'est pertinent?

\begin{figure}
\begin{center}
\null
\hfill
   \includegraphics[width=0.49\linewidth]{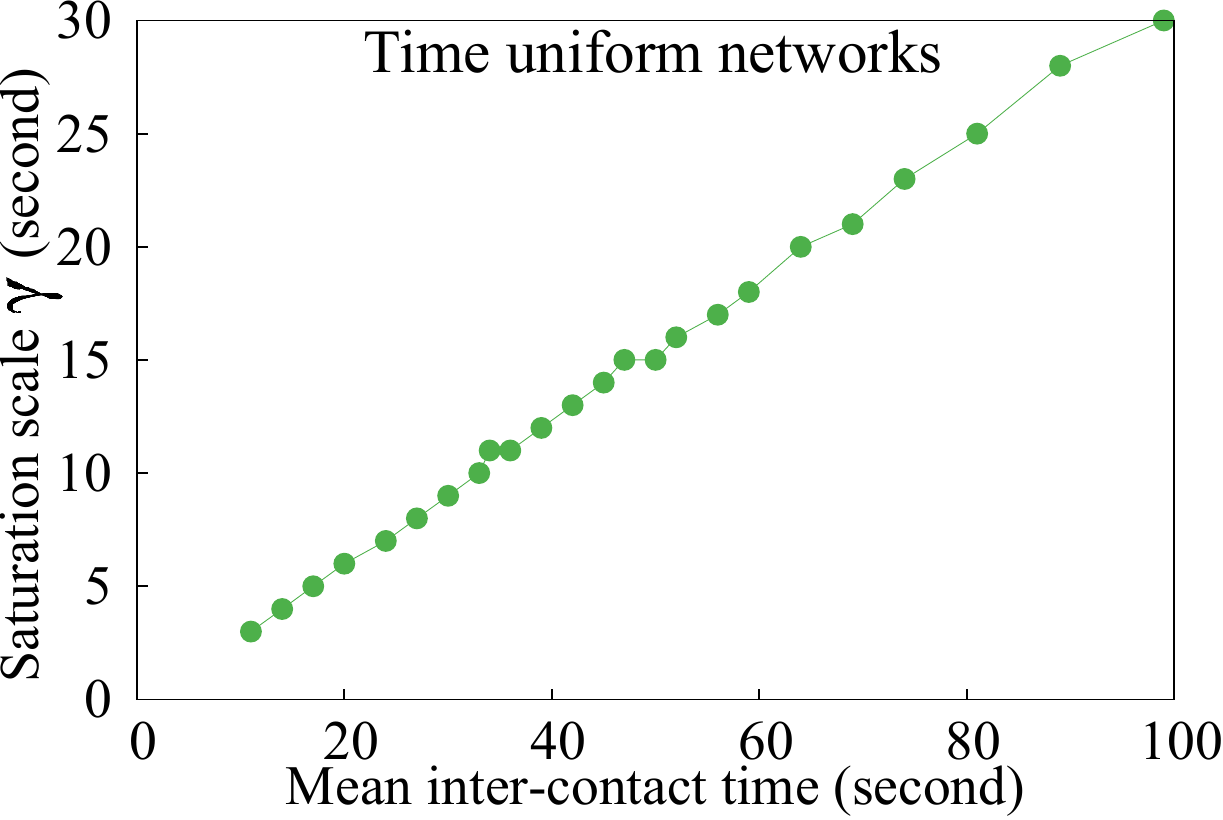}
\hfill
   \includegraphics[width=0.49\linewidth]{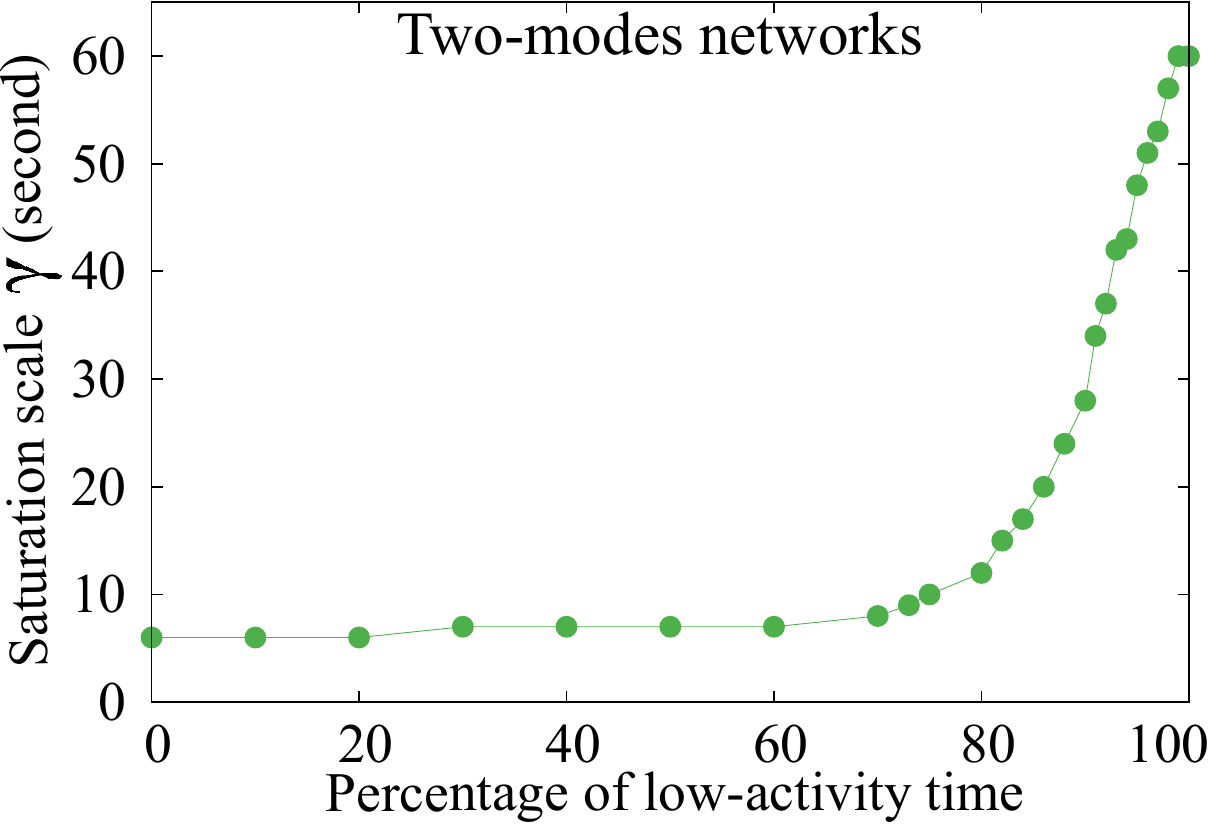}
\hfill
\null
\caption{Left: for time uniform networks, saturation scale (y-axis) in function of the mean inter-contact time of nodes (x-axis). Right: for two-mode networks, saturation scale (y-axis) in function of the percentage of low-activity time (x-axis).% Right: X axis : average (blue circles) and median (green square) of inter-contacts; Y axis : aggregation time obtained by the occupancy method.
}
\label{fig:day-night}
\end{center}
\end{figure}

We now investigate how the aggregation period returned by our method depends on the level of activity of the link streams considered, i.e. the number of links per node and per unit of time, and on the temporal heterogeneity of this activity. To this purpose, we use two kinds of synthetic dynamic networks, where the activity is uniformly distributed between all pairs of nodes. The first kind, called \emph{time uniform networks}, 
%the activity is also uniformly distributed along time, while in the second kind, the network has two modes of activity which alternate along time, one mode with a high level of activity and one mode with a low level activity.
is generated by assigning $N$ links ($N<<T$) to each pair of the $n=100$ nodes of the network and uniformly randomly choosing each of their timestamps between $0$ and $T=100\,000$s. We make the value of $N$ vary from $10$ to $100$ and for each of these values, we compute the aggregation period $\gamma$ returned by the occupancy method. Results are given in Figure~\ref{fig:day-night} left, which shows $\gamma$ as a function of the average inter-contact time of one node, that is $T/(N(n-1))$. For these time uniform networks, the aggregation period returned by the occupancy method is perfectly proportional to the average inter-contact time, showing that our method correctly takes into account the level of activity of the link stream.

%%%%%%%%%%%%%%%%
%%%%%%%%%%%%%%%%

However, most of the dynamic networks encountered in practice are far from being uniformly active over time. Many of them instead alternate periods of intense activity with periods of lower activity. In particular, this is the case for networks coming from human activities, such as the ones considered in Section~\ref{sec:general}, which often exhibit circadian rhythms. Then the question naturally arises to know how the saturation scale behaves according to this temporal heterogeneity. Does it simply make the average between the different levels of activity? Or does it favor one of them? To answer these questions we generate \emph{two-mode networks} that are built by $10$ alternations of one period of high activity and one period of low activity, which are time uniform networks with parameters $N_1,T_1$ and $N_2,T_2$ respectively. $N_1$, $N_2$ and the whole length $T=10(T_1+T_2)$ of study are fixed and we vary the ratio between $T_1$ and $T_2$.

Figure~\ref{fig:day-night} right gives the saturation scale $\gamma$ as a function of the percentage $\rho=T_2/(T_1+T_2)$ of low-activity time in the network. The curve goes from the value of $\gamma$ for the high-activity mode (for $\rho=0\%$) to the one, much larger, for the low-activity mode (for $\rho=100\%$). The plot shows that when the proportion of low activity varies from 0\% to 70-80\%, the saturation scale almost does not increase: it remains very close to the smaller value of the high-activity network, which preserves better the information contained in the original link stream. This is surprising as one would rather expect the saturation scale to be a compromise between its value for the low-activity periods and its value for the high-activity periods. This shows that in presence of heterogeneity of the activity along time, even with high-activity periods occupying only 30\% to 20\% of the time, the saturation scale returned by the occupancy method is respectful of this important part of the dynamics. Moreover, and importantly, the fact that the saturation scale does not linearly vary with respect to the percentage of low-activity time in the network shows that, for networks that are not time uniform (which is in particular the case of real-world networks), the saturation scale returned by the occupancy method does not only depend on the mean inter-contact time of nodes in the network (or equivalently on the frequency of links in the network).

When the proportion of low-activity time goes beyond 80\%, the aggregation period returned starts to increase until it reaches the value for the low-activity network when its proportion in time is 100\%. This seems natural as when the low-activity part takes most of the time of the dynamic network, it does not make sense to continue to study it with a scale which is suitable only for a marginal part of the time. Nevertheless, we note that the increase of $\gamma$ is progressive. For example for 90\% of low activity, the returned value is close to the arithmetic mean between the values for the two modes of the network. This shows that the returned aggregation does not forget too quickly the high activity part of the dynamics, which once again is a desirable feature of such a method.

%%%%%%%%%%%%%%%%%%%%%%%%%%%%%%%%%%%%%%%%%%%%%%%%%
%%%%%%%%%%%%%%%%%%%%%%%%%%%%%%%%%%%%%%%%%%%%%%%%%
\section{Detection of the more uniform distribution} \label{sec:selec}
%%%%%%%%%%%%%%%%%%%%%%%%%%%%%%%%%%%%%%%%%%%%%%%%%
%%%%%%%%%%%%%%%%%%%%%%%%%%%%%%%%%%%%%%%%%%%%%%%%%

%\marginpar{satisfactory/satisfying}

%%%%%%%%%%%%%%%%
\begin{figure}
   \includegraphics[width=0.49\textwidth]{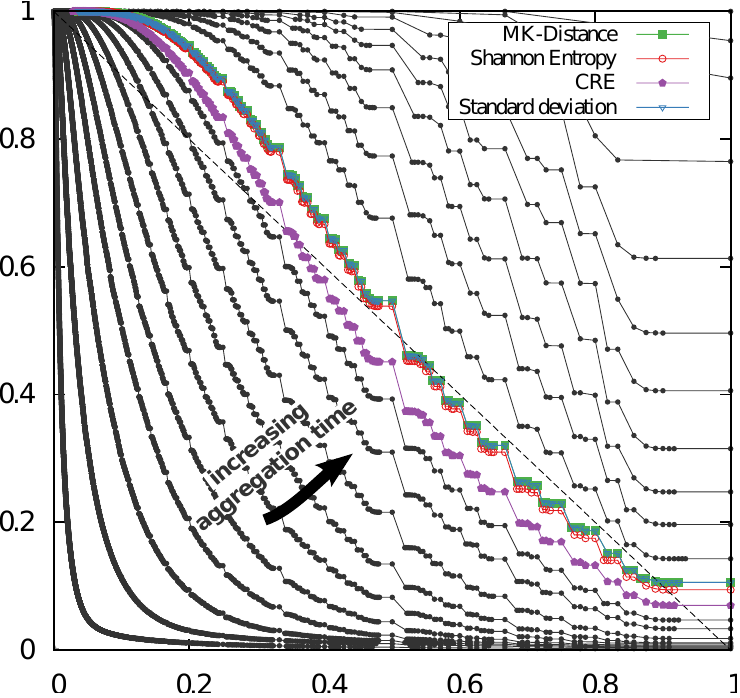}
   \includegraphics[width=0.49\textwidth]{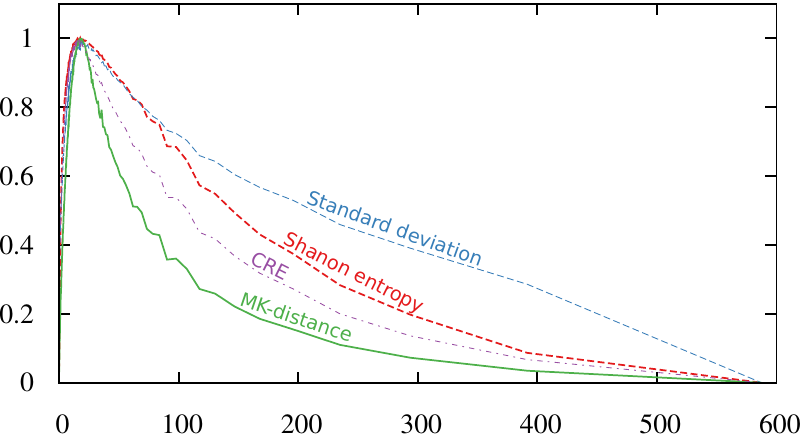}
\caption{Results of four methods for selection of the more uniformly spread distribution: M-K proximity, standard deviation, Shannon entropy with $10$ slots and cumulative residual entropy (CRE). The left plot shows the distributions selected by the maximum of each of the metrics and the right plot shows the variations (normalized to have maximum $1$) of each metric (y-axis) depending on the aggregation period $\Delta$ (x-axis).% The results for the variation coefficient are not depicted.%The maximum is reached for 18.7 hours.
}
\label{fig:compare_irvine}
\end{figure}
%%%%%%%%%%%%%%%%

%\marginpar{montrer methode coef de variation? non}
%\marginpar{zoom du max? non}
%\marginpar{different taille de slot pour shannon? non, texte only}

In this section we consider several methods for selecting the aggregation period $\gamma$ that gives the distribution of occupancy rates that is the more uniformly spread on $[0,1]$, and we study the dependence of $\gamma$ on the chosen selection method. Until now, all the results we gave were obtained by selecting the distribution which minimizes the M-K distance with the uniform density distribution on $[0,1]$. Of course, one may think of many other ways to select $\gamma$. This includes plotting the sets of distributions obtained when the aggregation period spans its entire range of variation, like on Figure~\ref{fig:method}, and selecting one distribution by visual mean. This empirical method is likely to give the most satisfactory results in practice and will therefore be preferred in many studies. However, here, we are interested only in quantitative methods of selection, for two reasons. Firstly, we want to provide a uniquely defined value which can be used as a reference for comparing the saturation scales of different dynamic networks. Secondly, we want our method to be fully automatic in order to be easily incorporable to any tool for analysis of dynamic networks.

In addition to the method based on the M-K distance, which we used until now, we now consider four other selection methods and compare their results. These four additional methods are based respectively on: standard deviation, variation coefficient, Shannon entropy and cumulative residual entropy. Figure~\ref{fig:compare_irvine} gives the results obtained when applying these methods on the Irvine data set. We now describe and analyze them one by one each before giving a global comparison of their respective results.

\paragraph{M-K distance with the uniform density distribution.} \begin{sloppypar}The Monge-Kantorovich distance is a way to measure the distance between two distributions of probability on the same support, here $[0,1]$. It is defined as the area comprised between the two inverse cumulative distributions of the probability distributions to be compared. Here, as we are looking for the distribution which is maximally spread over $[0,1]$, we compare each distribution with the uniform density distribution, which gives $dist_{M-K}(X)=\int_{[0,1]} |P(X>\lambda)-(1-\lambda)| d\lambda$, where $X$ is the random variable defined by the occupancy rate. Then, the aggregation period we select is the one for which the distribution of occupancy rates $X$ realizes the minimum of $dist_{M-K}(X)$. In order to get the desired distribution for the maximum of the measure, instead of the minimum, as for all the other measures we consider, we rather use the corresponding proximity measure defined as $1/2-dist_{M-K}(X)$, as $dist_{M-K}(X)$ is always less than $1/2$. This metric is the one we use throughout the article. It gives visually very satisfying results for all the data sets.\end{sloppypar}

\paragraph{Standard deviation.} This method selects the distribution having the maximum standard deviation $\sigma=\sqrt{E[(X-\mu)^2]}$, where $X$ is the random variable defined by occupancy rate and $\mu$ is its mean value. This is one of the most direct measure one can think of in order to compare the spread of distributions on support $[0,1]$. It gives very satisfactory results, comparable to the one obtained with the M-K distance. Nevertheless, it tends to select slightly higher aggregation period than the M-K distance, as the standard deviation is less penalized by the increasing of occupancy rates $1$, which is the maximal value in the distribution. Then, usually, the aggregation period selected by the M-K distance is visually a bit more satisfying. This is the reason why we prefer to present our methodology using the M-K distance, but the two metrics actually give comparable results.

\paragraph{Variation coefficient.} Another very natural method is the one that selects the distribution having the maximum variation coefficient $c_v=\sigma/\mu$, where $\mu$ is the mean and $\sigma$ the standard deviation of the distribution. Moreover, it could possibly correct the slight drawback of the standard  deviation pointed above, which tends to select a little bit higher value than would desire. Unfortunately, the method based on variation coefficient suffers from a much more severe limitation: it favors too much distributions having a small mean and therefore proposes only very short aggregation periods, or even not to aggregate at all. Among all the methods we tried, this is the only one which gives clearly unsatisfactory results to select the more spread distribution.

%\marginpar{reprendre le paragraphe sur shannon}

\paragraph{Shannon entropy.} In information theory, the Shannon entropy $H(X)=E[-ln(P(X))]$ of a random variable $X$ (here the occupancy rate) allows to measure the spread of distributions on a given fixed finite support. This means that in order to compare different distributions, the set of possible values taken by the distributions (the support) must be the same and must be finite. As for the measure based on the M-K distance, the distribution which maximizes the Shannon entropy is the one with uniform density on the considered support. The difficulty we face here to use this measure is that the supports of the distributions we want to compare are not the same: the set of possible values of the occupancy rate is different for each aggregation period. There are different ways to deal with this issue in order to compare all the distributions we obtain when varying the aggregation period. The first one is to artificially take one common support for all the distributions, the minimal such support being the union of the supports of all obtained distributions. Unfortunately, when applied with this support, the method always selects the distributions that effectively use the larger part of the support, that is those obtained for very short aggregation periods. As noted for the variation coefficient, this does not give satisfactory results.

Another possibility to solve this issue is to discretize the segment $[0,1]$ into $k$ slots of equal length and to compute for each distribution what is the probability that the value of the occupancy rate belongs to each slot. For example, when applied with $k=10$, this measures gives very satisfactory results. On the other hand, the returned aggregation period depends on the number of slots chosen. The results are sensibly different using $k=5$ or $k=20$. With smaller number of slots, the method tends to select higher aggregation periods, while with greater number of slots, like previously, it favors the distributions having greater original supports, which are those obtained for short aggregation periods. For $k=100$ this trend is already clearly marked: the value returned for $\gamma$ is less than half of the one returned with $k=10$ and visually, the distribution selected does not appear to be spread over $[0,1]$ in the best possible way among all the distributions obtained. Despite of this, we note that in the range of variation we consider for $k$, namely $[2,100]$, and for the data sets we use, the selection method based on the Shannon entropy properly determines the order of magnitude of the saturation scale $\gamma$ and even gives visually very satisfying results for values of $k$ between $5$ and $20$. Nevertheless, because of its sensitivity to the chosen $k$, we decided not to use this selection method. The next metric is another attempt to correct the difficulties arising from the use of the Shannon entropy.

\paragraph{Cumulative residual entropy (CRE).} This is a variation of the Shannon entropy that is able to compare distributions with the same \emph{infinite} support. It is suitable for our purpose as all the distributions we consider have support $[0,1]$. In this case, the cumulative residual entropy is defined as $\varepsilon(X)=-\int_{[0,1]} P(X>\lambda) \log(P(X>\lambda)) d\lambda$. As for the Shannon entropy, the maximum value of the CRE is reached for the uniform density on the considered support. It turns out that this selection method performs well on all the data sets we used. It gives aggregation periods close to the one obtained by the M-K distance, usually shorter. Visually the results are satisfying, even if on some example this method appears to favor a bit too much distributions with large supports. But this is only a slight effect and this method seems quite suitable for our needs, and theoretically well funded. The reason why we preferred the M-K distance with the uniform density distribution is that it is conceptually much simpler and gives as good results.\\

%\medskip

%18.7h  18.1h  17.5h  14.5h
Let us now compare the aggregation periods selected by the 5 methods above for the Irvine data set. First of all, let us note that all the selected aggregation periods are very close between $14.5$h and $18.7$h, except one of them, the one based on the variation coefficient. This method proposes an aggregation period of 1 second, which is the resolution of the timestamps, and the distribution it selects is very far from being uniformly spread on $[0,1]$. The variation coefficient method therefore appears not to be suitable for our purpose. On this data set, the distributions selected by the M-K distance and the standard deviation methods are exactly the same. They are the distribution obtained with an aggregation period of $18.7$h. The distribution selected by the method based on Shannon entropy with 10 slots of width 0.1 is almost indistinguishable from the previous one, it is the distribution obtained for an aggregation period of $18.1$h. These two distributions are indeed visually quite well spread on $[0,1]$ and therefore, these three methods give here very satisfying results, and they also did on other data-sets (not presented here).
The method based on the cumulative residual entropy selects the distribution obtained with an aggregation period of $14.5$h, which is slightly lower than the three previous method. In this case, this distribution is also visually very well spread on $[0,1]$ and then quite good for our purpose.

As a conclusion, except the method based on the variation coefficient, all the four other methods we considered appear to give satisfactory results on all the data sets we use. We chose the method based on the M-K distance because it is conceptually simple and it gives very satisfactory results. Beyond the slight differences between these methods, the fact that they all give very close values of $\gamma$ shows that each of them is sound and is appropriate to detect the aggregation period that maximally stretches the distribution of the occupancy rates in the interval $[0,1]$.

%\textcolor{red}{Above all, it demonstrates that the occupancy rate method does not consist in maximizing a specific function but catching a meaningful time scale.}

%%%%%%%%%%%%%%%%%%%%%%%%%%%%%%%%%%%%%%%%%%%%%%%%%
%%%%%%%%%%%%%%%%%%%%%%%%%%%%%%%%%%%%%%%%%%%%%%%%%
\section{Validation} \label{sec:valid}
%%%%%%%%%%%%%%%%%%%%%%%%%%%%%%%%%%%%%%%%%%%%%%%%%
%%%%%%%%%%%%%%%%%%%%%%%%%%%%%%%%%%%%%%%%%%%%%%%%%

%\begin{figure}
%\begin{center}
%    \includegraphics[width=0.8\linewidth]{./losses.pdf}
%\caption{The total length of the elongated paths over the total length of original paths.}
%\label{fig:loss-stretch}
%\end{center}
%\end{figure}

In this section we quantify the amount of information which is lost when one aggregates the network using a given period $\Delta$. This allows us to validate our approach by evaluating the loss obtained for $\Delta=\gamma$. Moreover, this provides tools to select more accurately an aggregation period, in the range preceding $\gamma$, that is suitable for representing a given link stream as a graph series.

The first measure of loss we use is the proportion of shortest transitions (minimal trips with two hops, cf. Definition~\ref{def:trans}) that lay entirely in one aggregation window. These are exactly the shortest transitions of the original link stream that do not exist anymore in the aggregated series of graphs: all the other minimal trips having their two hops, say $(a,b,t_1),(b,c,t_2)$, in two different aggregation windows, say indexed $t'_1$ and $t'_2$, still exist in the form $(a,b,t'_1),(b,c,t'_2)$ in the aggregated series. We chose this way of measuring the loss as the shortest transitions are the key units that capture the possibilities of propagation in the link streams. In other words, note that if all the shortest transitions of the link stream are conserved in the graph series (in the sense above), so are all the minimal trips, and therefore, the possibilities of propagation in the dynamic network are unchanged.

%There are a great number of paths on two hops in a time series, but only a limited number are particularly important for the property of propagation of the dynamic networks: the shortest paths in time. The shortest paths in time with two hops contain all the sequence of two hops that are necessary to communicate quickly in the dynamic network. In particular, note that if all the shortest path in time with two hops are preserved, then all the minimal trips in time are preserved and therefore the possibilities of communication in the dynamic network are unchanged. The loss of other paths on two hops does not threaten these possibilities. This is the reason why our measure of the loss is based only on shortest path in times with two hops: it simply counts the proportion of such paths which remains after aggregation.

\begin{figure}
\begin{center}
\null
\hfill
    \includegraphics[width=0.49\linewidth]{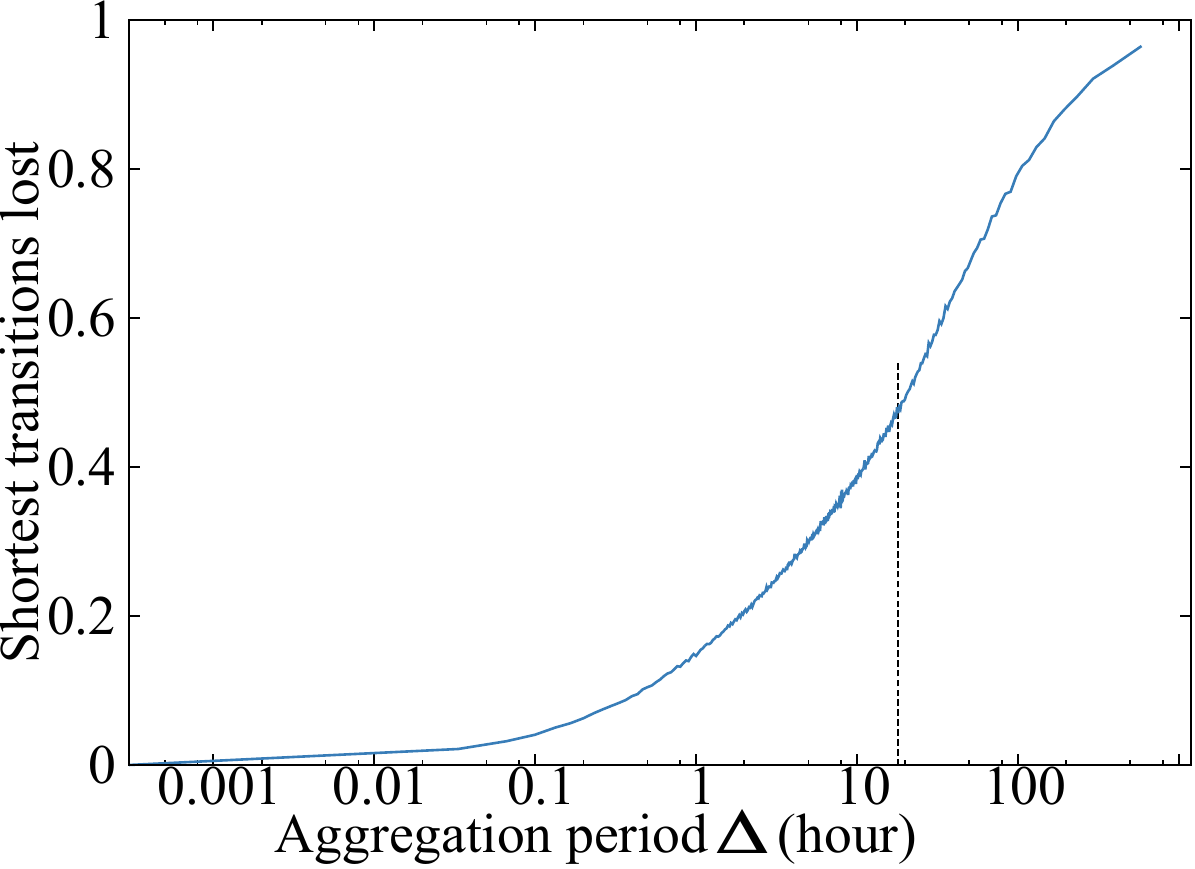}
\hfill
    \includegraphics[width=0.49\linewidth]{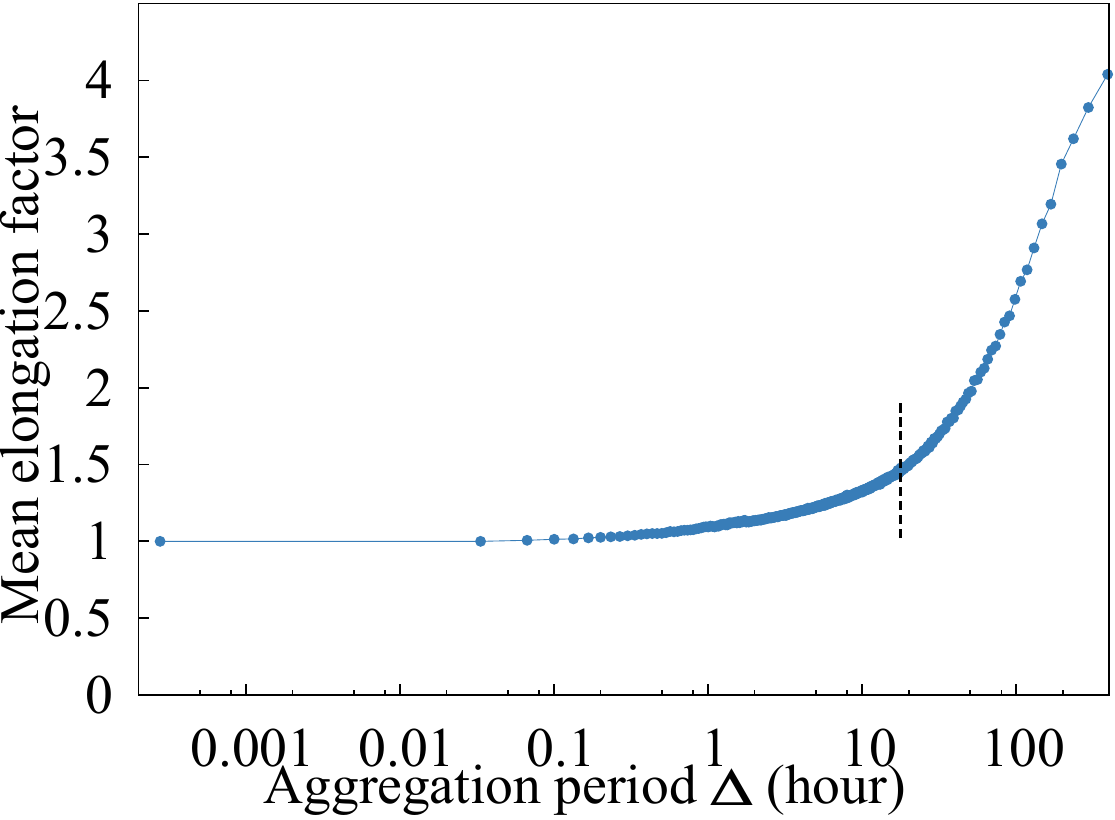}
\hfill
\null
\caption{Left: proportion of shortest transitions lost (y-axis) in the aggregated series $\cur{G}_{\Delta}$ according to the aggregation period $\Delta$ (x-axis). Right: mean elongation factor of minimal trips of $\cur{G}_{\Delta}$ (y-axis) according to $\Delta$ (x-axis).% (right) The percentage of the paths that are elongated with a percentage greater than N\%.
}
\label{fig:loss}
\end{center}
\end{figure}

Figure~\ref{fig:loss} left depicts the proportion of lost transitions as a function of the aggregation period $\Delta$, for the Irvine network. One can see that when the aggregation increases, starting from 1 second, the number of lost transitions first remains very low during several orders of magnitude, until an aggregation period of 0.5h where only 10\% have been lost. The main part of the loss ($80\%$) is concentrated on the range between 0.5h and 235h, i.e. a bit more than 2 orders of magnitude. The saturation scale $\gamma=18h$ returned by the occupancy method is in the beginning of this range, and in the middle in terms of order of magnitude. This shows that the occupancy method successfully detects the order of magnitude of the time scale from which the loss of information starts to be visible. For $\Delta=\gamma$, 48\% of the shortest transitions are lost. Therefore, one may prefer to limit further the range of aggregation periods used, for example one order of magnitude below $\gamma$.

%%%%%%%%%%%%%%%%
%%%%%%%%%%%%%%%%

On the other hand, it must be clear that the measure of the loss used above is rather pessimistic. Indeed, some of the shortest transitions of the original link stream that are lost can be replaced by some others slightly longer or occurring a bit later. This limits the actual impact of this loss on the possibilities of propagation in the aggregated series. As lost transitions can be replaced, the duration of a minimal trip that was using some of these transitions may be only slightly altered by their loss (or even not at all). For this reason, we also use a measure of loss which is based on the elongation of minimal trips in the aggregated series $\cur{G}_{\Delta}$ compared to the original link stream $\cur{L}$.

\begin{definition}[Elongation factor]
The \emph{elongation factor} $e_P$ of a minimal trip $P=(u,v,t_u,t_v)$ of $\cur{G}_{\Delta}$, with $t_u\neq t_v$, is defined as the ratio $$e_P=(t_v-t_u+1).\Delta/time_{\cur{L}}(P)$$ where $time_{\cur{L}}(P)=min\set{ t'_v-t'_u\ |\ (u,v,t'_u,t'_v)\textrm{ is a minimal trip of } \cur{L} \textrm{ and } t'_u,t'_v\in [(t_u-1).\Delta,t_v.\Delta]}$
\end{definition}

% \marginpar{verifier comment est calculee exactement l'elongation}
% ce serait pas mieux: la duree du plus court chemin dans la serie originale sur l'intervale [(t_u-1).\Delta,t_v.\Delta] avec contrainte de un saut par fenetre, divisee par la duree sur la meme fenetre toujours dans la serie originale et sans contrainte.
% note posterieure: c'est (très relié à) la question de casser les fenêtres de départ et d'arrivée, on l'avait envisagé ça en effet, mais peut-être pas dans la série???

Note that when $t_u\neq t_v$, we necessarily have $time_{\cur{L}}(P)\neq 0$. Therefore, the elongation factor is properly defined.
Figure~\ref{fig:loss} right gives the mean elongation factor (y-axis) of all minimal trips of the series aggregated with period $\Delta$ (x-axis), for the Irvine network. When $\Delta$ increases, the elongation factor of minimal trips first stays very close to $1$ during several orders of magnitude, before it suddenly raises when the aggregation period reaches values around the saturation scale $\gamma$. This shows that our method properly determines the scale at which the properties of propagation of the link streams start to be altered by aggregation.
For $\Delta=\gamma$, the mean elongation ratio of minimal trips is less than $1.5$, % besoin valeur numerique exacte : 1.47 ?
showing that despite the 48\% of shortest transitions lost, the propagation properties of the original link stream are not yet too drastically altered.

%%%%%%%%%%%%%%%%
%nouvelle courbe avec aggreg=gamma: le stretch des minimal trips (y-axis,lin) en fonction de leur duree initiale (x-axis,log)
%%%%%%%%%%%%%%%%

%%%%%%%%%%%%%%%%%%%%%%%%%%%%%%%%%%%%%%%%%%%%%%%%%
%%%%%%%%%%%%%%%%%%%%%%%%%%%%%%%%%%%%%%%%%%%%%%%%%
\section{Conclusion} \label{sec:conclu}
%%%%%%%%%%%%%%%%%%%%%%%%%%%%%%%%%%%%%%%%%%%%%%%%%
%%%%%%%%%%%%%%%%%%%%%%%%%%%%%%%%%%%%%%%%%%%%%%%%%

We showed that there exists a threshold, called the saturation scale $\gamma$, for the aggregation period of a link stream at which a qualitative change occurs in the way the network responds to aggregation. We showed that this change of behavior reveals an alteration of the properties of propagation of the dynamics, implying that dynamic networks should not be aggregated with a period larger than $\gamma$ to perform analyses that depend on these properties. In addition, we designed a fully automatic and parameter-free method to determine the value of $\gamma$ for an arbitrary link stream.

Our work open several perspectives to improve the method and broaden its field of application. The first of these perspectives is to extend the occupancy method to the case where links have a duration. The method presented in this article applies to both discrete and continuous time, to both undirected links and directed links, but it is able to deal only with links that are punctual events. However, in some contexts, the links of the dynamic network last during an interval of time (e.g. phone calls and physical contacts between individuals). Adapting the occupancy method to this case would be highly desirable. One particularly interesting way to do so would be to develop a notion of minimal trip that is specifically adapted to links that have a duration.

In Section~\ref{sec:synth}, we pointed out a nice behavior of the occupancy method in presence of temporal heterogeneity in the activity of the link stream processed: the aggregation scale $\gamma$ returned in this case gives more importance to the parts of the dynamics that have a high level of activity, even if they do not occupy the majority of the time. Nevertheless, if these periods are really too short, they will have only a limited impact on the value of $\gamma$. As a consequence, these highly active parts of the link stream, which are likely to contain a valuable information for the whole dynamics, may be smoothed out by the aggregation process. Avoiding this phenomenon by better taking into account the temporal heterogeneity of the activity of the link stream would constitute a key improvement. To this end, one could enhance the method so that it is able to separate the high activity periods from the lower activity periods and to determine an appropriate aggregation scale for each of these parts independently. Then one could decide either to aggregate the whole link stream at the shortest aggregation scale detected, which is the one that better preserves the information contained in it, or to partition the period of study and aggregate each part with a different length of window.

%Another key improvement of the method would be to better take into account the temporal heterogeneity of the activity of the link stream processed. 

%Another key improvement of the method would be to better take into account the temporal heterogeneity of the activity of the link stream processed. In Section~\ref{sec:synth}, we pointed out a nice behavior of the occupancy method to this regard: in presence of such an heterogeneity, the aggregation scale $\gamma$ returned by the method gives more importance to the parts of the dynamics that have a high level of activity, even if they do not occupy the majority of the time. Nevertheless, if these periods are too short, they will have only a limited impact on the value of $\gamma$. As a consequence, these highly active parts of the link stream, which may contain a key information for the whole dynamics, will be smoothed out by the aggregation process. In order to avoid this, one could enhance the method so that it is able to separate the high activity periods from the lower activity periods and to determine an appropriate aggregation scale for each of these parts independently. Then one could decide either to aggregate the whole link stream at the shortest aggregation scale detected, which is the one that better preserves the information contained in it, or to partition the period of study and aggregate each part with a different length of window.

%%%%%%%%%%%%%%%%%%%%%%%%%%%%%%%%%%%%%%%%%%%%%%%%%
%%%%%%%%%%%%%%%%%%%%%%%%%%%%%%%%%%%%%%%%%%%%%%%%%
\section*{Funding}
%%%%%%%%%%%%%%%%%%%%%%%%%%%%%%%%%%%%%%%%%%%%%%%%%
%%%%%%%%%%%%%%%%%%%%%%%%%%%%%%%%%%%%%%%%%%%%%%%%%

This work was supported by the French National Research Agency contract CODDDE [ANR-13-CORD-0017-01]. It was performed within the framework of the LABEX MILYON [ANR-10-LABX-0070] of Universit{\'e} de Lyon, within the program "Investissements d'Avenir" [ANR-11-IDEX-0007] operated by the French National Research Agency (ANR). The second author gratefully acknowledges the support from a grant from R{\'e}gion Rh{\^o}ne-Alpes and from the delegation program of CNRS.

%%%%%%%%%%%%%%%%%%%%%%%%%%%%%%%%%%%%%%%%%%%%%%%%%
%%%%%%%%%%%%%%%%%%%%%%%%%%%%%%%%%%%%%%%%%%%%%%%%%
\bibliographystyle{plain}
%\bibliographystyle{abbrv}
%{\small
\bibliography{aggregation,using-agreg,study-agreg}
%}
%%%%%%%%%%%%%%%%%%%%%%%%%%%%%%%%%%%%%%%%%%%%%%%%%
%%%%%%%%%%%%%%%%%%%%%%%%%%%%%%%%%%%%%%%%%%%%%%%%%

\end{document}